\documentclass[sigconf]{acmart} 
\AtBeginEnvironment{quote}{\par\singlespacing\small}

\AtBeginDocument{%
  \providecommand\BibTeX{{%
    \normalfont B\kern-0.5em{\scshape i\kern-0.25em b}\kern-0.8em\TeX}}}

\copyrightyear{2024}
\acmYear{2024}
\setcopyright{rightsretained}
\acmConference[CHI '24]{Proceedings of the CHI Conference on Human Factors in Computing Systems}{May 11--16, 2024}{Honolulu, HI, USA}
\acmBooktitle{Proceedings of the CHI Conference on Human Factors in Computing Systems (CHI '24), May 11--16, 2024, Honolulu, HI, USA}
\acmDOI{10.1145/3613904.3642211}
\acmISBN{979-8-4007-0330-0/24/05}

\begin{document}

\title{Investigating Use Cases of AI-Powered Scene Description Applications for Blind and Low Vision People}

\author{Ricardo Gonzalez}
\authornote{Both authors contributed equally to this research.}
\email{reg258@cornell.edu}
\author{Jazmin Collins}
\authornotemark[1]
\email{jc2884@cornell.edu}
\affiliation{%
  \institution{Cornell Tech}
  \city{New York}
  \state{New York}
  \country{USA}
}

\author{Cynthia Bennett}
\email{clbennett@google.com}
\affiliation{%
  \institution{Google}
  \city{New York}
  \state{New York}
  \country{USA}
}

\author{Shiri Azenkot}
\email{shiri.azenkot@cornell.edu}
\affiliation{%
  \institution{Cornell Tech}
  \city{New York}
  \state{New York}
  \country{USA}
}

\renewcommand{\shortauthors}{Gonzalez and Collins, et al.}

\begin{abstract}
"Scene description" applications that describe visual content in a photo are useful daily tools for blind and low vision (BLV) people. Researchers have studied their use, but they have only explored those that leverage remote sighted assistants; little is known about applications that use AI to generate their descriptions. Thus, to investigate their use cases, we conducted a two-week diary study where 16 BLV participants used an AI-powered scene description application we designed. Through their diary entries and follow-up interviews, users shared their information goals and assessments of the visual descriptions they received. We analyzed the entries and found frequent use cases, such as identifying visual features of known objects, and surprising ones, such as avoiding contact with dangerous objects. We also found users scored the descriptions relatively low on average, 2.76 out of 5 (SD=1.49) for satisfaction and 2.43 out of 4 (SD=1.16) for trust, showing that descriptions still need significant improvements to deliver satisfying and trustworthy experiences. We discuss future opportunities for AI as it becomes a more powerful accessibility tool for BLV users.
\end{abstract}
\nopagebreak



\begin{teaserfigure}
\includegraphics[width=\textwidth]{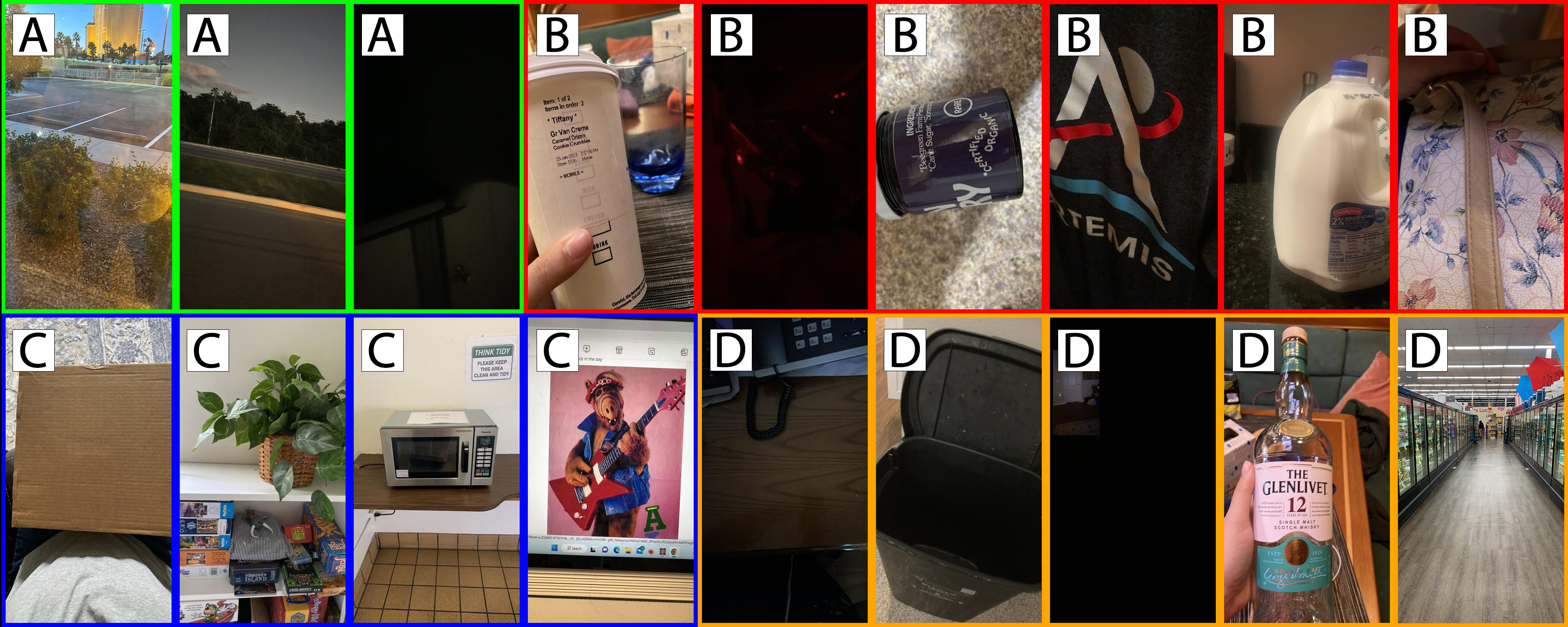}
\caption{ During the diary study, participants submitted entries with different goals. These examples represent four of the most common goals: (A) Getting a scene description, (B) identifying features of objects, (C) obtaining the identity of a subject in the scene, and (D) learning about the application. }
\Description{18 images randomly sampled from the diary study. These examples represent four of the most common goals: The first three images are examples of (A) Getting a scene description,  the next six images are examples of (B) identifying features of objects, the next four images are examples of (C) obtaining the identity of a subject in the scene, and the last five are examples of (D) learning about the application. }
\label{fig:teaser}
\end{teaserfigure}


\maketitle

\section{Introduction} \label{Introduction}

Blind and low vision (BLV) people face many daily challenges in mundane tasks. Seemingly simple questions can become frustrating ordeals. For example, when picking something to wear, a BLV person may wonder: “Which of these blouses is the one with pink flowers on it?” When trying to eat healthy, she might inquire: “Does this protein bar have soy?” And when looking after her son, she might need to know: “Did my toddler throw food on the living room rug?” In recent years, \textit{visual interpretation} applications have become available to help BLV people answer such questions. They allow users to submit a photo or video to a remote service that returns information about its visual contents.

Researchers have studied the use of applications that describe visual information extensively. The most significant work on this topic explored the use of VizWiz and Aira, visual description applications that allow users to send a photo and question to a remote sighted assistant. Brady et al. \cite{brady-visual-challenges-everyday-lives} analyzed over 40,000 VizWiz submissions and categorized user questions. They found that some of the most common types of questions involved identifying objects, describing things, and reading text. Lee et al. \cite{lee2018conversations} studied the use of Aira, interviewing  BLV users as well as sighted assistants. Similar to Brady et al.’s findings, they found that users sought descriptions of things and asked questions to support navigation. Aira users also wanted information about social interactions. These studies shed light on use cases for visual description applications, but they only examined use of applications with remote sighted assistants. A growing number of visual description applications are powered by AI rather than humans (e.g., Seeing AI \cite{seeingAI} and BeMyAi \cite{be-my-ai})–they leverage computer vision and natural language processing rather than sighted assistants. Thus, at this time, despite the burst of interest in AI, little is known about how these AI-powered applications are being used.

Although AI and human-powered applications both address visual access needs, BLV people may use them in different ways. People’s sense of trust, perceptions of privacy, and feelings of independence are likely to differ when requesting information from AI rather than a human; this may affect when, where, and how they use an application. Prior work has found that users have both over trusted and been skeptical of  AI-generated captions \cite{image-explorer-guo} and it is unknown how such varying levels of trust affect use of an application. Another study found that human assistance can detract from people’s sense of independence \cite{collins2023guide}, which may lead to increased use of AI-powered applications compared to their human-powered counterparts.

Despite the growing prominence of AI, little work has studied the use of AI-powered scene description applications (at the time of this study, a more common deployment of AI was to describe scenes captured in images; thus, we refer to the AI-powered applications mentioned here as scene description applications). Kupferstein et al. \cite{kupferstein2020understanding} briefly described a pilot diary study with four users of SeeingAI. Their preliminary findings show that participants used the application to seek private information and learn about objects they could not touch (e.g., an object behind a glass wall). Interestingly, participants were satisfied with only 43.0\% of interpretations. This initial exploration raised critical questions: In what other use cases do BLV people use AI-powered scene description applications? And why were participants not satisfied with the majority of interpretations? In short, there is presently a gap in our understanding of the use of AI-powered scene description applications.

We addressed this gap by building on Kupferstein et al.’s work with a rigorous investigation of an AI-powered scene description application. We posed the following research questions:

\begin{itemize}
    \item What are BLV people’s use cases for AI-powered scene description applications?
    \item How well are BLV people's needs addressed in these use cases? 
\end{itemize}

To answer these questions, we conducted a two-week diary study with 16 BLV participants. During this period, participants used a scene description application that we developed for this study. Working similarly to SeeingAI, the application described images using a state-of-the-art image analysis model developed by Microsoft \cite{microsoft-azure-ai, SeeingAI-Azure-API} and produced descriptions for photos submitted by the user. After providing a description, the application presented the participant with questions about the information they received, the user’s goal, and their use context, which all constituted a diary entry. At the end of the diary period, we conducted interviews with participants where we reviewed their entries and their use of the application.

Based on an extensive analysis of the diary entries, we categorize and describe use cases for the application in terms of user goal, photo content, location, and photo quality. Common user goals included identifying objects and visual features about them, understanding one’s surroundings, and reading text. We also uncovered some surprising goals, such as avoiding disgusting or dangerous objects, checking for others' presence, and resolving visual disputes with BLV peers. Participants rated their level of trust and satisfaction with each description, yielding a mean trust score of 2.43 out of 4 (SD=1.16) and a mean satisfaction score of 2.76 out of 5 (SD=1.49).  Interestingly, satisfaction and trust scores were not consistently aligned with description accuracy because users were sometimes able to infer information from poor or partially correct descriptions.

We discuss opportunities for future AI-powered scene description applications, arguing that designers should leverage BLV users’ contextual knowledge and experience deciphering imperfect descriptions. We also discuss the importance of exploring AI-powered scene description separately from human-powered visual interpretation and offer future directions for scene description applications.

In summary, we contribute an extensive discussion of use cases for AI-powered scene description applications, along with user assessments of AI-generated descriptions. We hope our work will support researchers and developers as they improve AI-powered accessibility tools, from developing the AI models to designing the applications themselves. In light of the rapid advancement of AI, our work comes at a pivotal moment where we can make AI-powered accessibility tools useful and accessible as they emerge.

\section{Related Work} \label{Related Work}

Our work contributes to research on visual descriptions for BLV people and their daily visual descriptive needs. We draw from two related threads: systems that provide visual information about one’s surroundings, and systems that describe images encountered on the Web. We refer to the former category as visual interpretation (human-powered) and scene description (AI-powered) systems, and present work on both approaches (Sections \ref{Human-Powered Visual Interpretation Systems} and \ref{AI-Powered Scene Description Systems}). We then provide an overview of the latter category, which we refer to as image description systems.

\subsection{Human-Powered Visual Interpretation Systems} \label{Human-Powered Visual Interpretation Systems}

Researchers have developed human-powered visual interpretation systems that connect BLV users to sighted human assistants. BLV users can use these systems to submit queries with photos to a sighted assistant who then returns an answer within several seconds \cite{bigham-vizwiz, naik2023contextvqa, lasecki2014increasing}.

One of the first systems of this kind was VizWiz, a crowd-powered visual question answering (VQA) mobile application deployed by researchers \cite{bigham-vizwiz}. VizWiz received thousands of interpretation requests, yielding a dataset of photos and questions that has been studied extensively \cite{brady-visual-challenges-everyday-lives, gurari2018vizwiz, gurari2019vizwiz, tseng2022vizwiz}. Brady et al. examined submissions from over 5,000 VizWiz users to reveal common visual challenges that blind people encounter in daily life. They created a taxonomy of visual questions and identified situations where the application failed to meet their needs. They found several categories of common questions which were grouped into four high-level categories: identification, description, reading, and unanswerable questions \cite{brady-visual-challenges-everyday-lives}. Following Brady et al., Gurari et al. found that several mainstream VQA models, despite performing well on the MSCOCO dataset \cite{lin2014microsoft} and other popular benchmark VQA datasets \cite{Thomee_2016, ImageNet, silberman2012indoor}, performed poorly on the VizWiz dataset. Gurari et al. presented this study to encourage the development of CV algorithms that work with BLV users’ photos \cite{gurari2018vizwiz}, due to the challenges they present to VQA.

More recent visual interpretation services enable real-time conversations between sighted assistants and users. Unlike VizWiz, sighted assistants can access a video feed of the user’s environment, rather than still images. Formative research gathered user and assistants insights  \cite{avila-be-my-eyes-survey, lee2018conversations, lannan2019virtual}. For example, Lee et al. investigated the uses of Aira, a well-known visual interpretation tool, throught two interview studies with sighted assistants and BLV users. In these studies Lee et al. found four use cases for human-powered live visual interpretation: scene description, navigation, task performance (performing tasks), and social engagement \cite{lee2018conversations, lee2020emerging}. Other researchers also investigated ways to improve information quality by studying both trained and untrained visual interpreters’ ability to collaborate to assist one   user \cite{xie-helping-helpers, xie-pair-volunteers}. Their subsequent taxonomy of visual interpretation tasks denote how labor can be divided among multiple agents \cite{xie-pair-volunteers}.

The above work demonstrates that human-powered systems are useful and enhance BLV people’s quality of life. Investigations into tools like Aira have explored how AI-powered assistance can enhance human support \cite{sooyeon-opportunities-human-ai-collab}, but there lacks an examination into the uses of AI-only visual interpretation services. Our work addresses this gap by identifying BLV users' AI-powered scene description use cases in daily life.

\subsection{AI-Powered Scene Description Systems} \label{AI-Powered Scene Description Systems}

Researchers have examined AI-powered systems that use CV models to generate visual descriptions of images BLV users upload. 

Some researchers have performed studies to record how well AI-powered scene descriptions work in daily life. For example, Smith and Smith performed a one-day autoethnographic diary study to explore the authors’ experiences and perspectives on the usefulness of “readily available [not specified],” AI-powered technologies. In one author’s diary, a visually-impaired researcher, she stated she felt the description application she relied on was most useful when it provided relevant information, and most frustrating when it provided nothing relevant, forcing her to rely on sighted assistance for clarification \cite{smithandsmith2021artificial}.

Past work has also evaluated commercialized AI-powered tools, such as SeeingAI \cite{granquist2021evaluation, kupferstein2020understanding}. Closest to our work, Kupferstein et al \cite{kupferstein2020understanding} presented preliminary results from a pilot diary study exploring four participants’ use of SeeingAI. Participants used SeeingAI when seeking information they wanted to keep private, and participants were frequently unsatisfied with resulting descriptions (only 43\% of reactions were positive). This extended abstract revealed that AI can supply information BLV people do not feel comfortable requesting from humans, but it is not clear as to why and when.

Overall, prior research has found that AI-powered applications can address some visual needs of BLV users. However, evaluations have been limited to anecdotal accounts and brief studies. An in-depth approach is necessary to fully understand user experiences and the uses of AI-powered scene description. Our work addresses this with a rigorous diary study that captures natural use of an AI-powered scene description application.

\subsection{Image Description}

Many researchers have examined descriptions of visual Web content, which are often called “alt-text” since they are communicated through the respective HTML attribute. Image description systems are similar to visual interpretation systems; however, this research mostly concerns images taken by sighted people that are consumed by BLV people browsing the Web.

Past work has explored factors contributing to an image description’s effectiveness in providing relevant information to BLV users \cite{gleason-failed-twitter-accessibility, muehlbradt-and-kane-what's-in-an-alt-tag, stangl-what-pvi-want-in-image-descriptions, stangl-beyond-one-size-fits-all-descriptions}. Stangl et al. conducted interviews with 28 BLV screen reader users, discussing their thoughts on alt-text they received across different types of websites (e.g., news sites, shopping sites, blogs). They found users’ preferences for type and specificity of details in descriptions varied depending on the image’s digital context \cite{stangl-what-pvi-want-in-image-descriptions}. Researchers have also conducted studies to evaluate the effectiveness of image descriptions \cite{gleason2019memes, kreiss-contexual-metrics, CIDER, SPICE}. For example, Kreiss et al. used participants’ ratings to evaluate the quality of alt-text descriptions in images found in online articles. They recruited 74 participants through Amazon Mechanical Turk, both sighted and BLV, to rate alt-text for images included in an article. They found these ratings could be used as indicators of description quality, and to assess how contextually pertinent information impacted descriptions’ perceived quality \cite{kreiss-contexual-metrics}.

In summary, no standard has emerged for evaluating descriptions of visual content. Despite this, it is important to establish a methodology for rigorous evaluation that assesses how well a visual description addresses BLV users’ needs. Thus, we took lessons learned from the space of image description and evaluation and applied them to our work. We implemented quality metrics with user-based scores for satisfaction and trust in descriptions and researcher scores on a description’s accuracy. The open-ended field in each diary entry and our follow-up interviews allowed us to collect pertinent yet context-specific information   to understand their use cases for AI-powered visual interpretation and the reasoning behind their scores.

\section{Methods} \label{Methods}

To examine use cases for scene description applications, we conducted a diary study with 16 BLV users. We developed an application that enabled participants to take photos of their surroundings and receive descriptions of scenes captured in the photos. Participants used this application throughout a two-week period, providing information about their experience every time they received a description.

\subsection{Participants}

\begin{table*}
  \caption{Participant demographics. The description of vision is self-reported by participants.}
  \label{table: participants}
  \begin{tabular}{p{0.8cm}p{1cm}p{1cm}p{1cm}p{1.6cm}p{3.7cm}p{1.5cm}p{4.4cm}}
    \toprule
   \textbf{PID} & \textbf{Age} & \textbf{Gender} & \textbf{Vision} & \textbf{From Birth} & \textbf{Vision Description} & \textbf{Use SeeingAI}  & \textbf{Assistive Technology}                                                            \\
   \midrule
1      & 24           & Female          & Low Vision & N/A      & Read with right eye. See colors & Yes &  VoiceOver, magnification, and contrast  \\
   \hline 
2      & 67           & Male            & Blind      & No      & Light perception. In extreme conditions identify color. & Yes & VoiceOver \\
\hline 
3   & 57           & Male            & Blind      & Yes      & Read with magnification with left eye only.  & Yes & CCTV to magnify print, talking thermometer and scale, SeeingAI, and VoiceOver      \\
\hline 
4                       & 34           & Female          & Blind      & Yes    & Quite near-sighted.  Sensitivity to light. Able to read with magnification  300\%.  & Yes     & CCTV to magnify print,  magnification, VoiceOver, and JAWS                                         \\
\hline 
5                       & 55           & Male            & Blind      & No     & Only light perception on  the periphery  & Yes  & JAWS,NVDA, VoiceOver, and Voice assistants  \\                                                           
\hline 
6                       & 52           & Female          & Blind      & Yes    & No Vision  & Yes   & VoiceOver, and Voice assistants                                                           \\
\hline 
7                       & 39           & Female          & Low Vision & No     & Light, color and shape perception. Varies a lot   & Yes             & ZoomText, and CCTV to magnify                        \\
\hline 
8                       & 44           & Male            & Blind      & Yes    & No Vision   & Yes           & VoiceOver                               \\
\hline 
9                       & 37           & Female          & Blind      & No     & No Vision   & Yes           & JAWS, VoiceOver, and Braille Display                              \\
\hline 
10                      & 39           & Male            & Blind      & Yes    & Minimal light perception.   & Yes               & VoiceOver                 \\
\hline 
11                      & 36           & Female          & Blind      & Yes    & No Vision   & Yes           & VoiceOver, JAWS, and Braille Display                             \\
\hline 
12                      & 32           & Female          & Low Vision & No     & Can see large objects, and colors contrast. Can not read                          & Yes   & JAWS, and VoiceOver      \\
\hline 
13                      & 39           & Female          & Blind      & No     & No Vision   & Yes           & JAWS, and VoiceOver                                                 \\
\hline 
14                      & 67           & Female          & Low Vision & No     &  Read with magnification.   & Yes     & JAWS, Braille Display, and magnification                                \\
\hline 
15                      & 47           & Female          & Blind      & No     & Light perception, color perception, and shapes perception.   & Yes & VoiceOver, and JAWS                             \\
\hline 
16                      & 59           & Female          & Blind      & No     & No Vision & Yes  & VoiceOver, Voice Assistants, and Braille Display\\
\bottomrule
\end{tabular}
\end{table*}
We recruited US-based adult participants through mailing lists from the National Federation of the Blind. We had three inclusion criteria: participants had to self-report non-correctable visual impairment, be over the age of 18, and be willing to use their iPhone, running iOS 9 or higher (application requirement), during the study.

We had 16 participants in total: five participants were men and 11 were women, with ages ranging from 24 to 67 years old (mean=45.5, SD=12.8). Twelve participants identified as blind and 4 identified as low vision. In Table 1, we list these and other self-reported demographics in detail. Participants were compensated with a \$60 Amazon gift card upon completion of the study.

\subsection{Procedure} \label{procedure}

The study consisted of a training session, a two-week application use period, and a follow-up interview. The training session was conducted over a video-conferencing platform and lasted 60 to 90 minutes. During the session, we asked participants about their demographics and experience with AI-powered scene description applications. We then trained them on the use of the study application. This included installing the application on the participant’s mobile device, explaining how to use the application, and submitting a test diary entry (which we did not include in our analysis).

During the diary period, we asked participants to use the application whenever they sought visual information about their surroundings, with a minimum requirement to submit one photo per day. However, we understood that the application, being designed only for scene description, would likely not fulfill every visual need participants had in two weeks (e.g., participants could not read text with the application). Thus, participants were permitted to use other tools to meet their daily needs during the study.

A diary entry included the photo taken, the returned description, and the user’s responses to a brief questionnaire, which was incorporated into the flow of the application. Based on Carroll’s definition of a use case \cite{carroll2003making}, our questionnaire collected information about the user’s goal and context with the following questions:

\begin{itemize}
    \item Where were you when you took the photo?
We instructed participants to briefly describe where they used the application (e.g., at my office, at my friend’s apartment, inside the airport, etc.).
    \item What information did you want to get from the description? 
We instructed participants to provide the reason for which they requested visual assistance (e.g., I wanted to know what was outside my window, I was looking for a pair of jeans on my shelf, etc.).
    \item How satisfied were you with the description?
Answer choices included:
Very dissatisfied, Somewhat dissatisfied, Not satisfied nor dissatisfied, Somewhat satisfied, and Very satisfied
    \item How much do you trust the description? 
Answer choices included:
Not at all, A little, Somewhat, and To a great extent.
    \item Add any additional comments.
\end{itemize}

We chose different scales for trust and satisfaction, with different numbers and types of options (e.g., “negative” or “positive” for trust, and “negative” to “neutral” to “positive” for satisfaction). For the trust scale, we aimed to capture whether participants were willing to take action based on a description. According to Dumouchel, trust is an action where a person gives something power over themselves, rather than a feeling motivating them; in other words, trust is a choice of action or inaction based on another agent \cite{dumouchel2005trust}. Thus, we chose a 4-point Likert scale for trust because we wanted participants to express their willingness to take action, from “Not at all” (would not take action) to “To a great extent” (would very likely take action). For the satisfaction scale, we aimed to capture whether participants’ expectations were met. Thus, we chose a 5-point scale from “very dissatisfied” to “very satisfied”, with a neutral value in the middle.

After participants completed the diary portion of the study, we conducted semi-structured interviews that lasted 50 to 90 minutes. The interviews were divided into two parts: (1) clarifying use cases from the diary study, and (2) reflecting on the usefulness of the descriptions. During the first part, we inquired about specific diary entries, asking questions like, “What did you do with the information the app provided?” or “Did you want to get a specific piece of information about the background?”. In the second part, we delved into participants’ broader experiences with the application, posing questions such as, “In the context of AI-powered applications, what does trust in an application mean to you?” or “What does satisfaction in an application mean to you?” We asked follow-up questions as needed. Finally, we asked the following to gain insight into participants’ overall evaluation of the application:

\begin{itemize}
    \item Overall, how trustworthy do you consider the descriptions provided by the application?	
Answer choices included:
Extremely untrustworthy, Untrustworthy, Somewhat untrustworthy, Neither trustworthy nor untrustworthy, Somewhat trustworthy, Trustworthy, and Extremely trustworthy.
    \item Overall, how satisfied are you with the descriptions provided by the application?
Answer choices included:
Extremely dissatisfied, Dissatisfied, Somewhat dissatisfied, Neither dissatisfied nor satisfied, Somewhat satisfied, Satisfied, and Extremely satisfied.
\end{itemize}
\begin{figure*}
\includegraphics[width=0.9\textwidth]{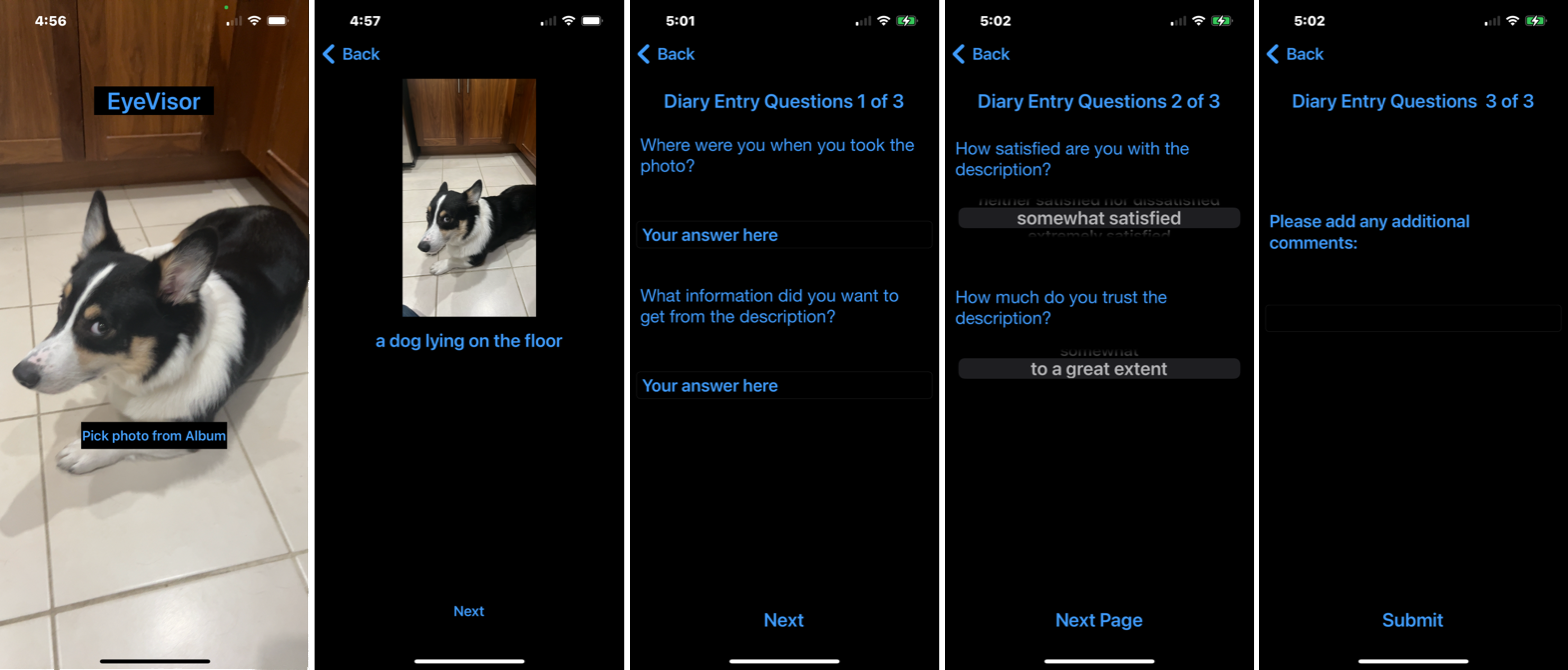}
\caption{\label{fig: EyeVisor} The scene description application we used to collect data. Screenshots show the flow of using the application and submitting a diary entry. It includes five screens: the photo submission, the photo description, and three diary entry question screens (see \ref{procedure} for questions in questionnaire). The interface was designed to group similar questions, while minimizing the number of elements on each screen.
}
\Description{ The scene description application we used to collect data. The screenshots show the flow of using the application and submitting a diary entry. It includes five screens: the photo submission, the photo description, and three diary entry question screens. The photo submission screen has a label that says Eyevisor, a button that says pick photo from album, and a dog in the background that is being captured with the camera. The visual interpretation screen contains a preview of the image captured, the visual interpretation result that says “a dog lying on the floor”, and two buttons “next”, and “back”. The first question screen contained the questions: (1) Where were you when you took the photo and (2) What information did you want to get from the description. The second question screen contained the questions: (3) How satisfied are you with the description, and (4) How much do you trust the description. Finally, in the third questions screen participants could enter additional comments. The interface was designed to group similar questions, while minimizing the number of elements on each screen.
}
\end{figure*}

\subsection{Scene Description Application}

We wanted to create an accessible diary study experience where participants could respond to questions immediately after receiving their photo description \cite{rieman-diary-study}. Thus, we developed a custom application to collect our study data. To ensure ecological validity, our application required comparable functionality to commercial applications like Microsoft’s SeeingAI and Google's Lookout (commonly used by BLV users). To that end, we used Microsoft’s Azure AI Vision image description API since it “exposes the latest deep learning algorithms” \cite{APIDocumentation, microsoft-azure-ai} and is also used by SeeingAI \cite{SeeingAI-Azure-API}. This API offers image analysis capable of detecting, classifying objects, and generating descriptions, based on a Microsoft dataset of more than “10,000 [image analysis] concepts” \cite{APIFeatures, microsoft-azure-image-analysis-documentation}. Employing this API, our application emulated SeeingAI’s “scene channel,” meaning it could describe environments, identify objects, identify text (but not read it), describe humans, animals, and plants and their actions, and also describe relationships between them.

The flow of the application was simple: when participants launched the application, they could take a photo or upload one from their device’s gallery. Upon submission of a photo, the application presented a results screen that included the description. If satisfied with the results, participants could press the “next” button to load the diary questionnaire (see figure \ref{fig: EyeVisor}). There were three questionnaire screens, and they included five questions in total. In the last questionnaire screen, participants finalized their submission by pressing the “submit” button.

\begin{table}
  \caption{ Number of diary entries submitted by participants.}
  \label{table: diaryentries}
  \begin{tabular}{p{3cm}p{4cm}}
    \toprule
    \textbf{Participant ID} & \textbf{Number of Diary Entries}                                                        \\
    \midrule
    \multicolumn{1}{c}{1} & \multicolumn{1}{c}{21} \\
    \multicolumn{1}{c}{2} & \multicolumn{1}{c}{32} \\
    \multicolumn{1}{c}{3} & \multicolumn{1}{c}{26} \\
    \multicolumn{1}{c}{4} & \multicolumn{1}{c}{15} \\
    \multicolumn{1}{c}{5} & \multicolumn{1}{c}{21} \\
    \multicolumn{1}{c}{6} & \multicolumn{1}{c}{23} \\
    \multicolumn{1}{c}{7} & \multicolumn{1}{c}{20} \\
    \multicolumn{1}{c}{8} & \multicolumn{1}{c}{20} \\
    \multicolumn{1}{c}{9} & \multicolumn{1}{c}{20} \\
    \multicolumn{1}{c}{10} & \multicolumn{1}{c}{39} \\
    \multicolumn{1}{c}{11} & \multicolumn{1}{c}{23} \\
    \multicolumn{1}{c}{12} & \multicolumn{1}{c}{23} \\
    \multicolumn{1}{c}{13} & \multicolumn{1}{c}{13} \\
    \multicolumn{1}{c}{14} & \multicolumn{1}{c}{21} \\
    \multicolumn{1}{c}{15} & \multicolumn{1}{c}{18} \\
    \multicolumn{1}{c}{16} & \multicolumn{1}{c}{19} \\
  \bottomrule
\end{tabular}
\end{table}

\subsection{Qualitative Analysis} \label{Qualitative Analysis}

We collected a total of 354 diary entries from our 16 participants. On average, each participant submitted 22 entries. The participant with the highest number of submitted entries was 39, and the participant with the lowest number was 13 (see table \ref{table: diaryentries}). Out of 354 entries, 22 (6.2\%) did not return interpretations due to connectivity or technical issues, so they were discarded from our analysis. Out of the remaining 332, 16 were removed because they were taken by each participant once at the beginning of the study to test the application. The final number of entries analyzed was 316.

The goal of our qualitative analysis was to capture the full context of each diary entry to identify participants’ use cases for scene descriptions. We supplemented our analysis by incorporating user’s perspectives from the follow-up interviews.

Our analysis involved two parts: (1) an inductive open coding process of the diary entries and the follow-up interviews \cite{saldana2021coding}, and (2) affinity diagramming sessions to group the resultant codes \cite{rex-hartson-ux-methods-affinity}. For the diary entries, we coded participants’ questionnaire responses and the content of their submitted photos. For the follow-up interviews, the open coding process organically revealed connections with diary codes, and also revealed underlying themes behind the use cases for scene description applications. Finally, in the affinity diagramming sessions, we grouped codes to create the themes we share in Findings (see Section \ref{Findings}).

For the diary entries, we derived codes based on the participant’s goal in the entry and context in which they used the application. We defined \textit{User Goal} as the reason for taking the photo (e.g., identifying a specific object, determining an object’s color), drawn from participant’s self-reported goals (diary entry response to “What did you want to get from the description?”), photo content, and statements in supporting interviews. Most diary entries had one goal, but some had multiple. To code context, we examined all information relating to the “what was captured” and “where it happened” of each diary entry. The “what” included the content of the photo and photo properties based on visual qualities that may impact the API’s performance (e.g., lighting conditions, photo clarity). The “where” included the self-reported location where the photo was taken and the image’s setting (e.g., indoors, outdoors, private, public). Two researchers coded 30 diary entries individually, then came together to discuss discrepancies and generated a codebook. Afterwards, remaining entries were split between researchers and codes were added or clarified as needed.

\begin{table*}
  \caption{Shows the five coded user goal levels, definitions, and examples applied under each level.}
  \label{table: goals}
  \begin{tabular}{p{2.5cm}p{5cm}p{4.5cm}p{3cm}}
    \toprule
    \textbf{User Goal} & \textbf{Definition} & \textbf{User Goal Examples} & \textbf{Image Example}                                                        \\
    \midrule
    Describe Scene   & Gathering visual information about related subjects that form a scene          & ``Is there a mountain in the background?''
\newline
``Was my room messy or not?''
     &  \begin{tabular}[c]{@{}c@{}} 
{
\includegraphics[width=60pt, height=90pt]{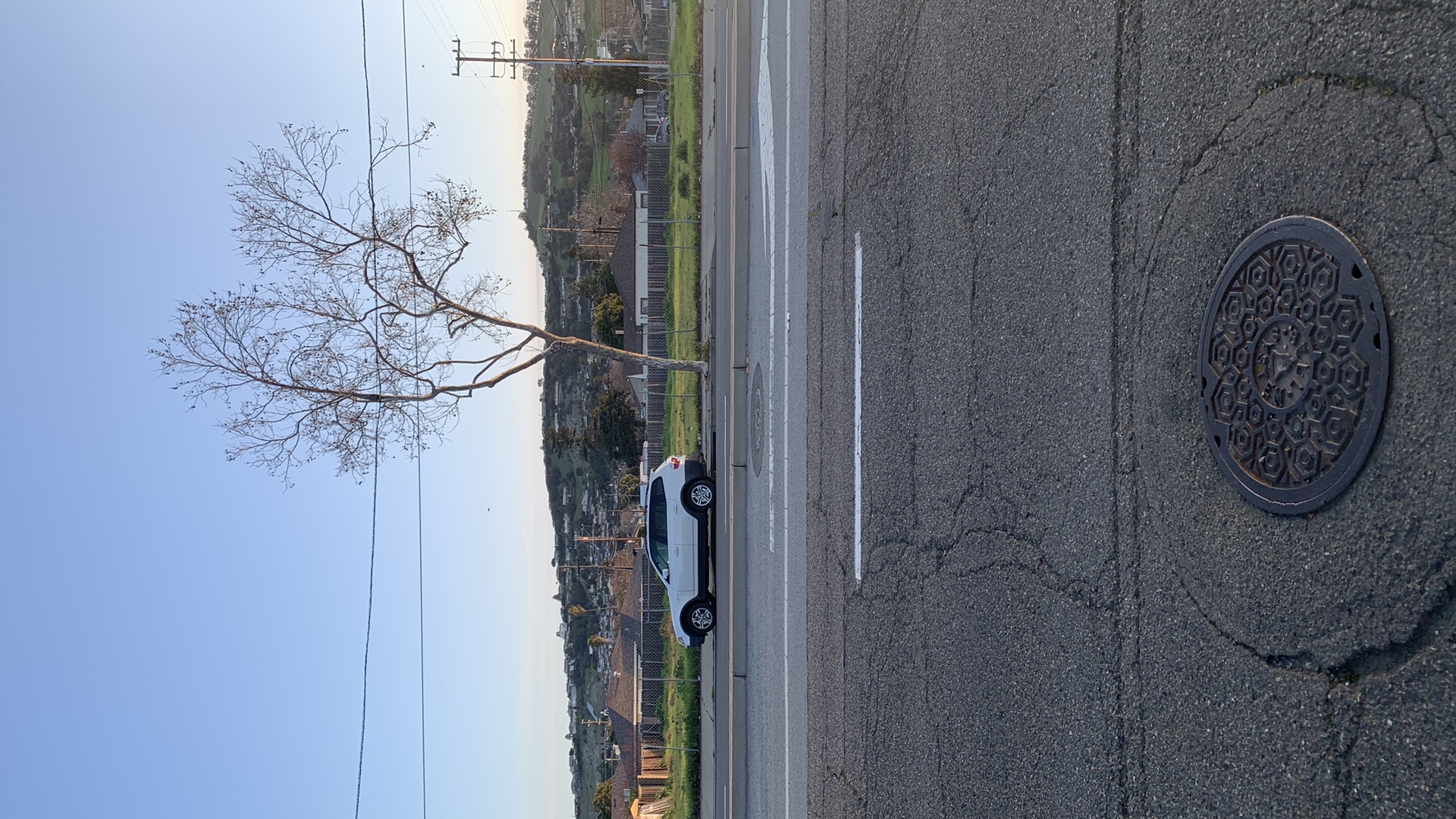}
\Description{This photo represents an example of an entry coded for user goal “describe scene”. Participant asked “Is there a mountain in the background?”. A photo of a street with a white care, a tree with no leaves in the background, a mound and clear blue sky. 
}
}

 \end{tabular} \\
     Learn Application   & Learning how to use the application, testing how well the application performed, or becoming familiar with the application output   & ``I was testing to see if it could describe the pattern on this shirt.'' ``I wanted to use the app to help me frame a photo of my dog.''
     &  \begin{tabular}[c]{@{}c@{}} {\includegraphics[width=60pt, height=90pt]{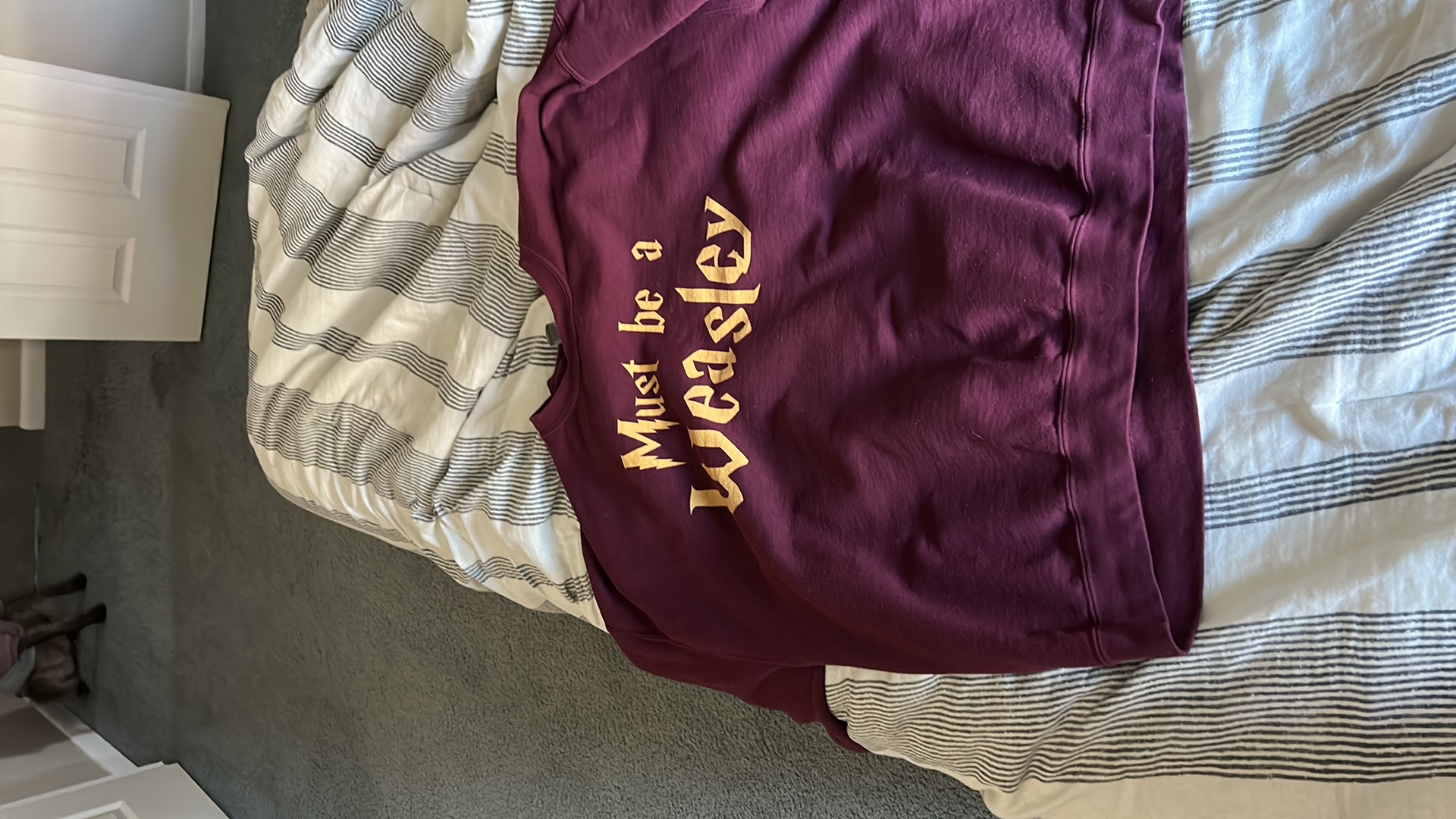}
\Description{This photo represents an example of an entry coded for “learn application”. Participant mentioned in the diary entry “I was testing to see if it could describe the pattern on this shirt”. A photo of a burgundy shirt with yellow text that says: “Must be a Weasley” on a bed.
}}
 \end{tabular} \\
     Identify Subject   & Identifying a specific subject when the user has no clear working knowledge of the subject  & ``What was the image my friend just tweeted?'' ``What item am I holding?''
     &  \begin{tabular}[c]{@{}c@{}} {\includegraphics[width=60pt, height=90pt]{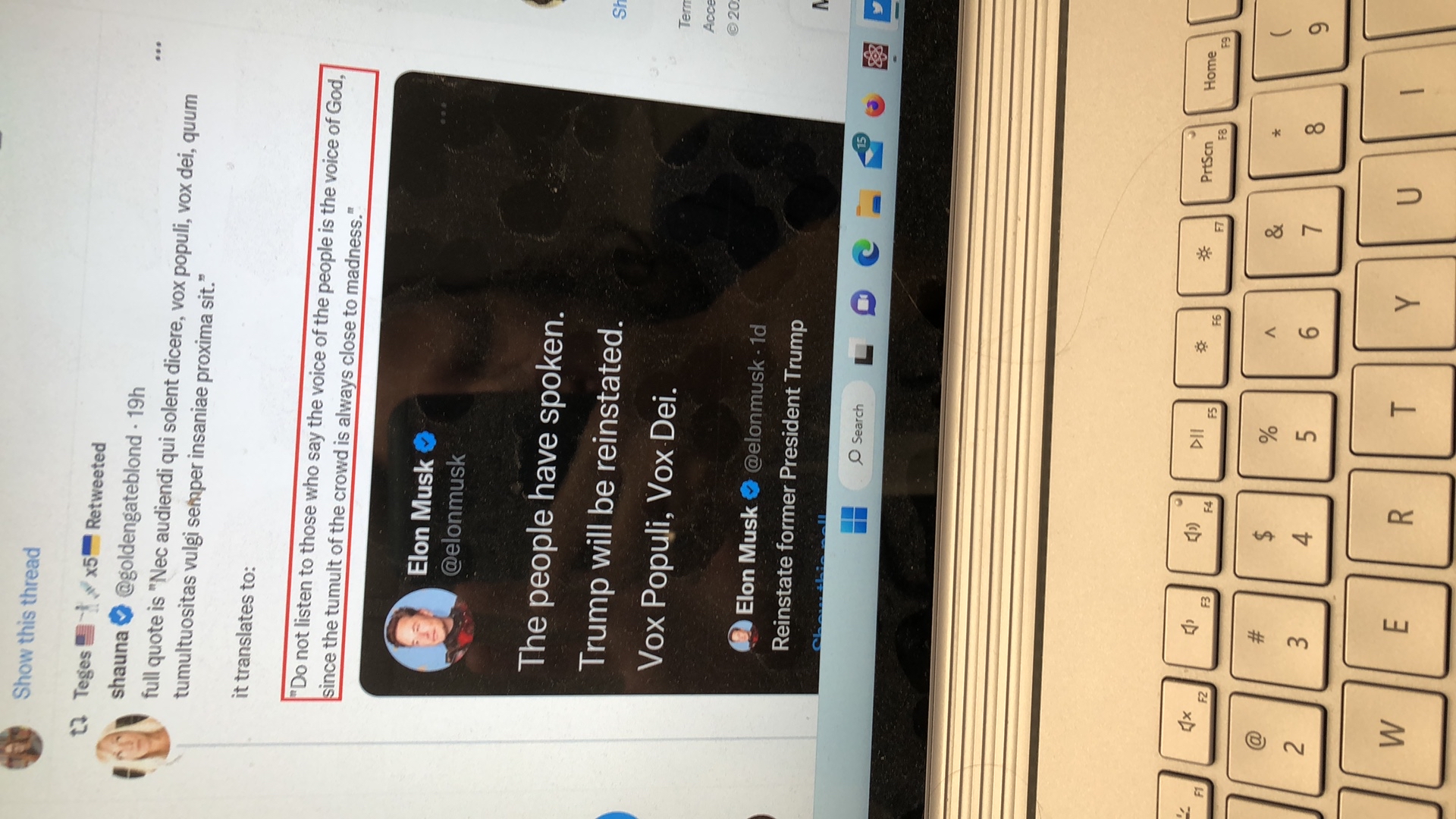}
\Description{This photo represents an example of an entry coded  for user goal “identify subject”. Participant asked “What was the image my friend just tweeted?”. A photo of a  macbook laptop displaying a Tweet of elon musk. It says “The people have spoken. Trump will be reinstated. Box Populi, Vox Dei”.
}}
 \end{tabular} \\
     Identify Feature           & Identifying features of a subject such as color, size, state (on/off, full/empty), and so on, when a user already has a working knowledge of the subject's identity          & ``Is the lamp in my room on or off?'' 
     \newline ``What is the logo on this baseball glove?''
     &  \begin{tabular}[c]{@{}c@{}} \rotatebox{0}{\includegraphics[width=60pt, height=90pt]{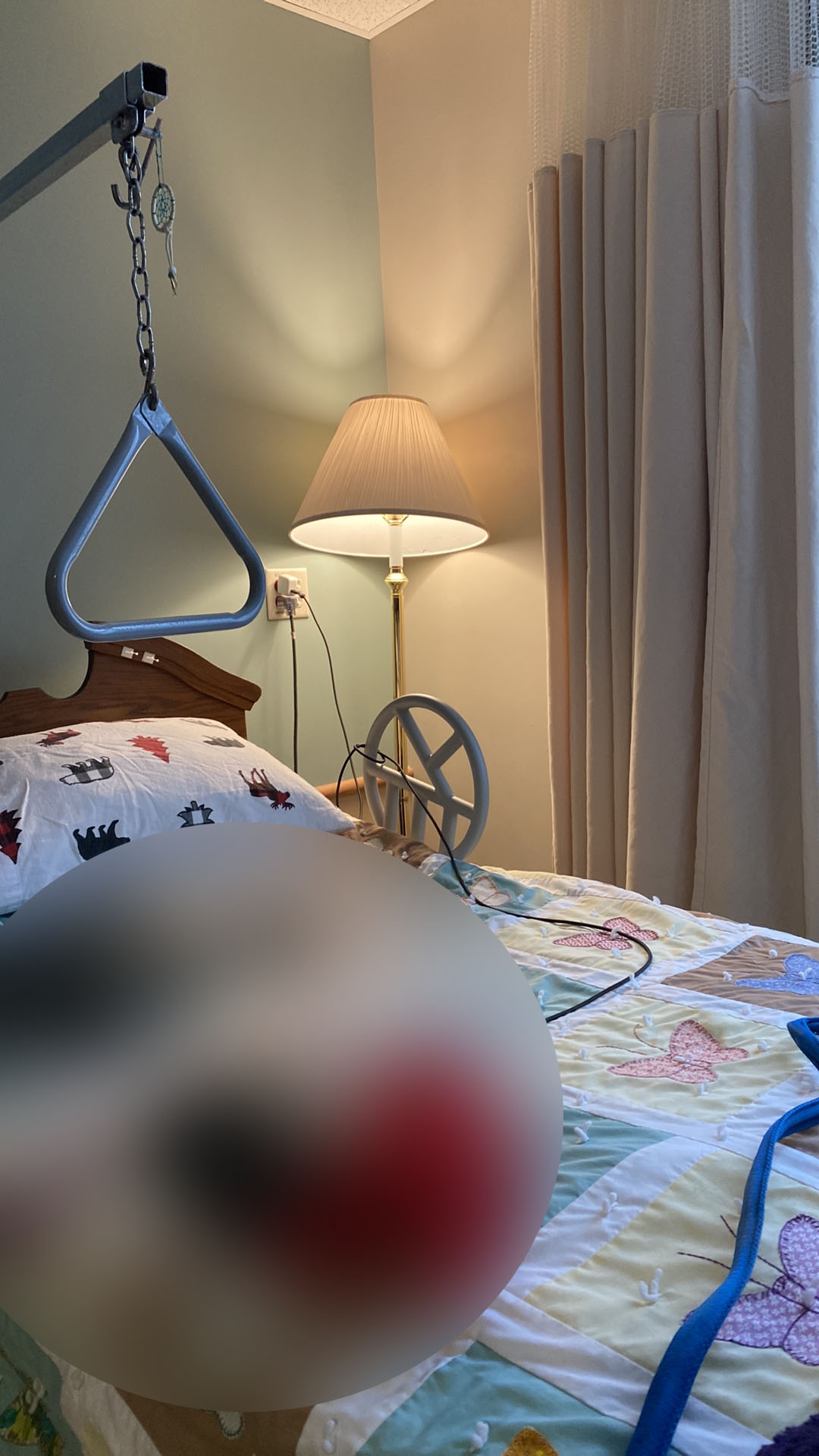}
\Description{This photo represents an example of an entry coded  for user goal “identify features”. Participant asked “Is the lamp in my room on or off?”. A photo of a bedroom. There is a bed with colorful sheets and pillow, a device to help users with mobility disabilities stand up and a lamp that is turned on.
}}
 \end{tabular} \\
     No Specific Goal       & Goal was either unclear to the coders, or the user mentioned they had no goal in mind &``I just wanted a basic description''
     \newline
``People''
     &  \begin{tabular}[c]{@{}c@{}} \rotatebox{0}{\includegraphics[width=60pt, height=90pt]{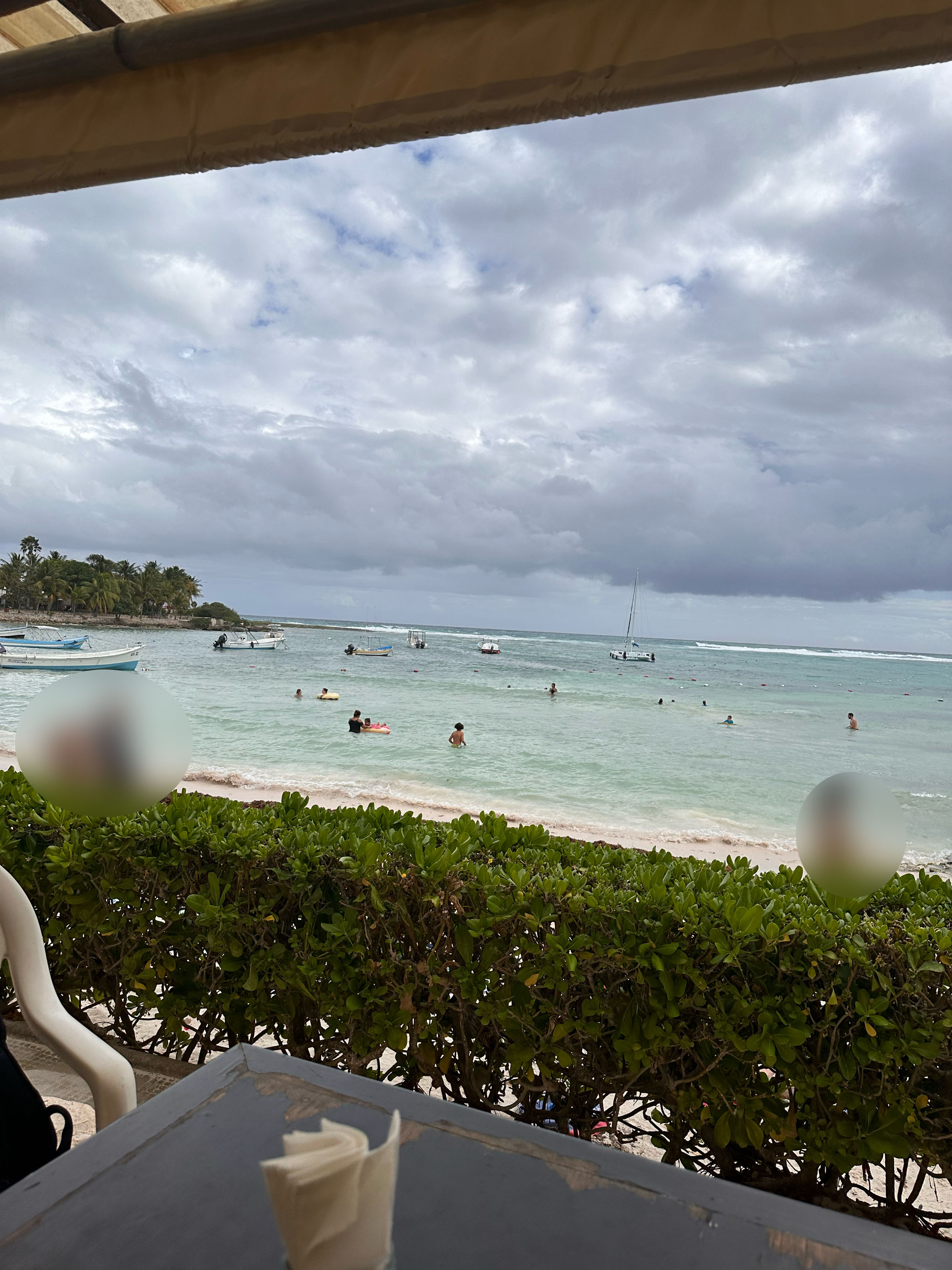}\Description{This photo represents an example of an entry coded  for user goal “no specific goal”. Participant mentioned in the diary entry “I just wanted a basic description”. A photo of a a table and a green hedge with a clean beach in the background. The sky looks cloudy. 
}}
 \end{tabular} \\    
  \bottomrule
\end{tabular}
\end{table*}
\begin{table*}
  \caption{Part 1 of coded locations. Shows four coded location levels, definitions, and examples applied under each level.}
  \label{table: locations}
   \begin{tabular}{p{4cm} p{4cm} p{3cm}}
    \toprule
    \centering \textbf{Location} & \centering \textbf{Location Examples} & \textbf{Image Example}                                                        \\
    \midrule
    Used in educational or spiritual sites    & Schools, churches, and \newline other religious institutions
     &  \begin{tabular}[c]{@{}p{1cm}@{}} 
\rotatebox{90}{\includegraphics[width=60pt, height=90pt]{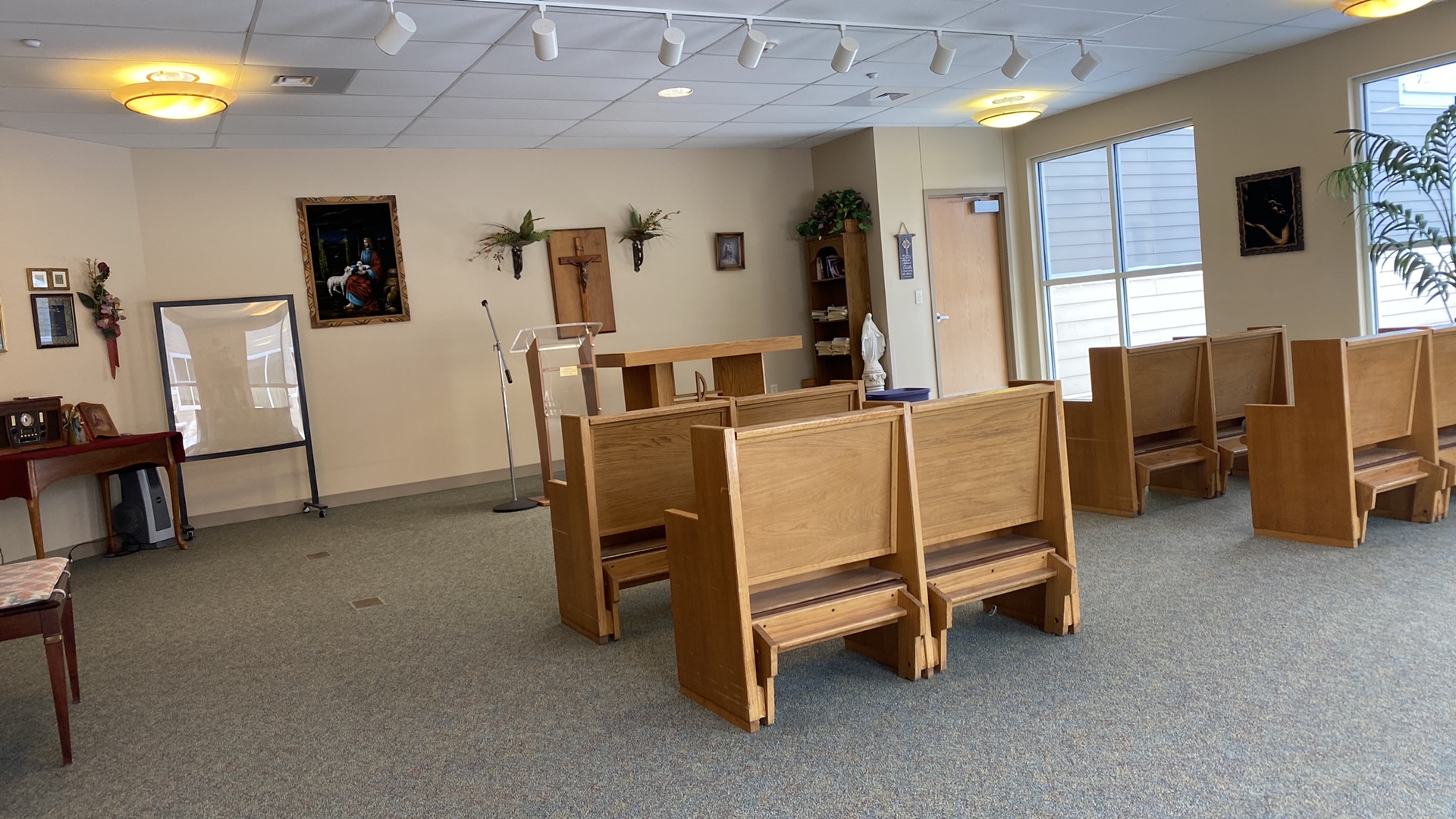}\Description{This photo represents an example of an entry coded for location “Used in educational or spiritual sites”. Participant was located in a chapel. A photo of an empty chapel with wooden pews.
}}
 \end{tabular} \\
    Used in recreational or leisure areas   & Restaurants, parks, and historical sites
     &  \begin{tabular}[c]{@{}p{1cm}@{}} 
\rotatebox{0}{\includegraphics[width=60pt, height=90pt]{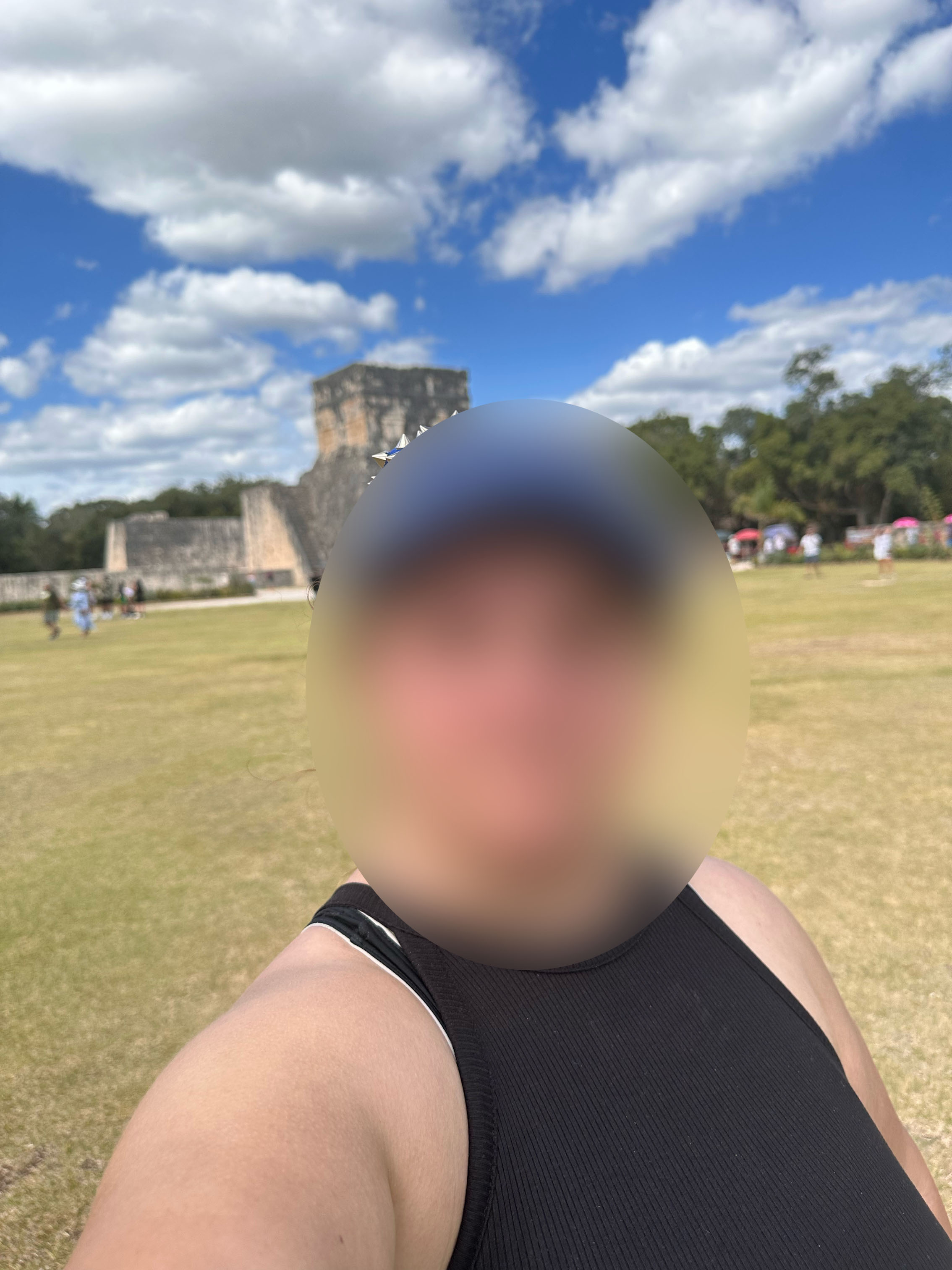}\Description{This photo represents an example of an entry coded for location “Used in recreational or leisure areas”. Participant was located in a historical site. A photo of a woman taking a selfie in an open field with an aztec pyramid in the background.
}}
 \end{tabular} \\
     Used in healthcare places & Hospitals, dentist offices, and ophthalmologist offices   
     &  \begin{tabular}[c]{@{}p{1cm}@{}} 
\rotatebox{0}{\includegraphics[width=60pt, height=90pt]{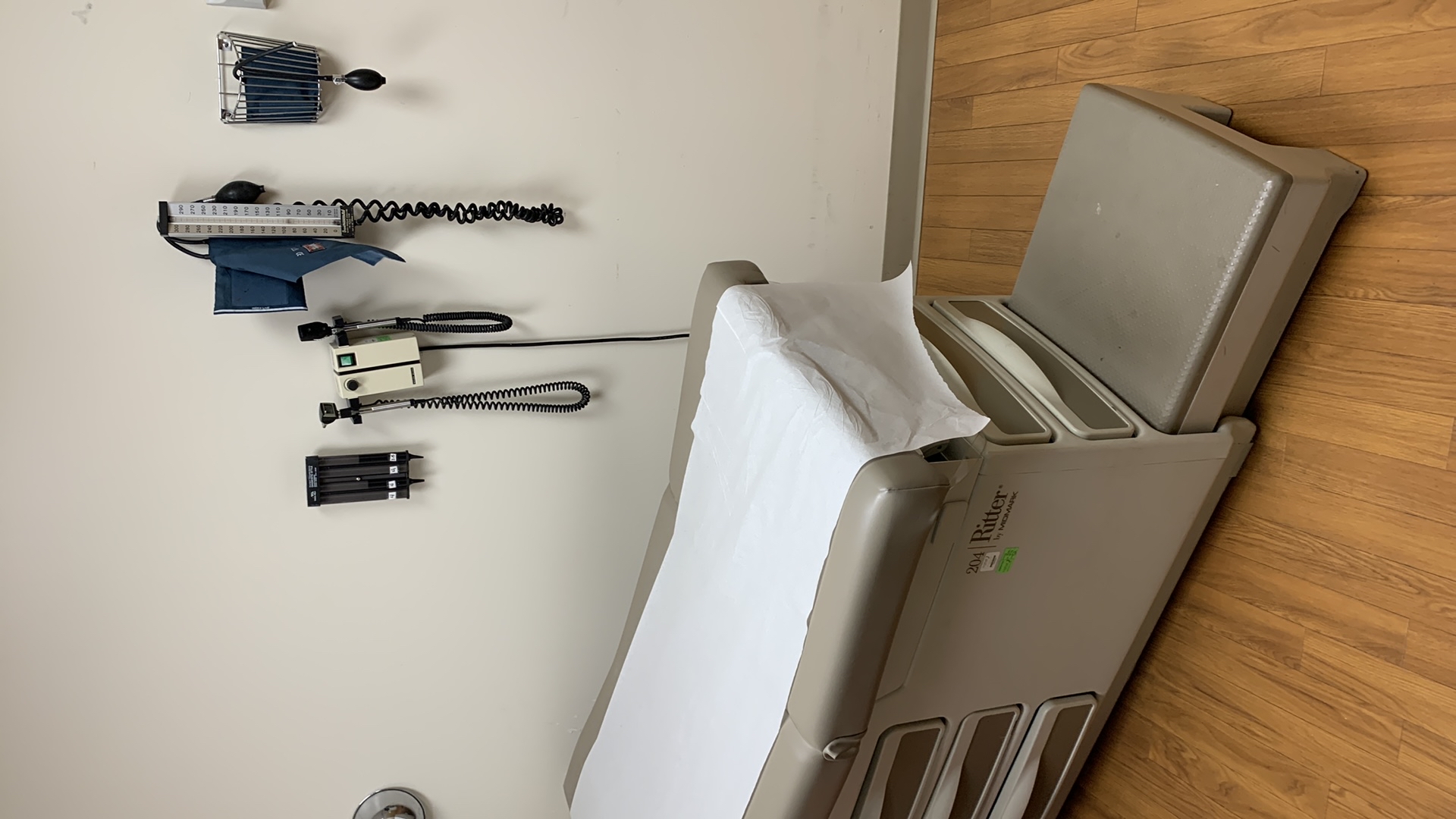}\Description{
This photo represents an example of an entry coded for location “Used in healthcare sites”. Participant was located in an optometrist office. A photo of a group of medical equipments for general health checks and an examination table.
}}
 \end{tabular} \\
     Used in travel services sites  & Bus stops, train stations, and airports
     &  \begin{tabular}[c]{@{}p{1cm}@{}} 
\rotatebox{0}{\includegraphics[width=60pt, height=80pt]{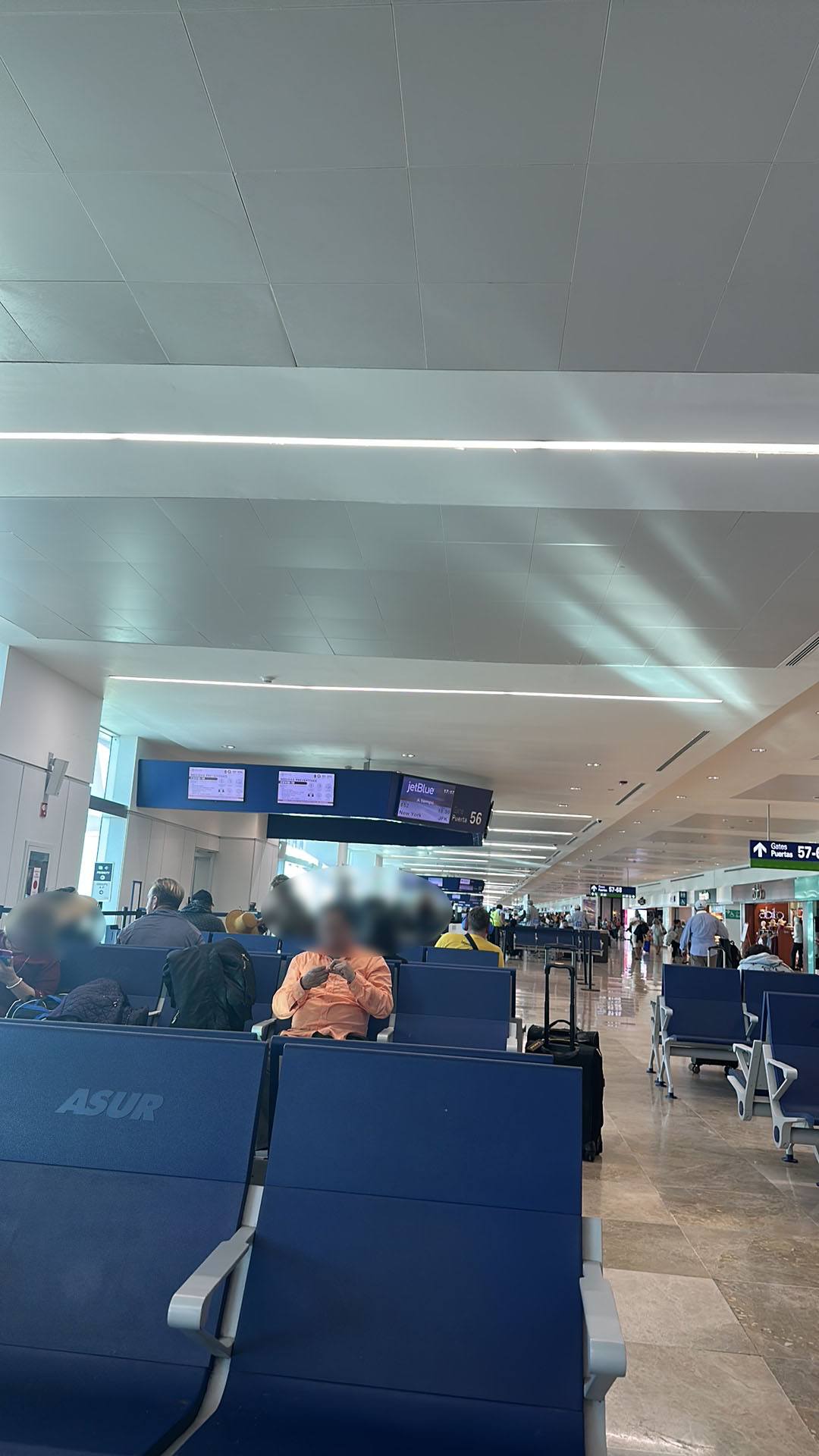}\Description{This photo represents an example of an entry coded for location “Used in travel service sites”. Participant was located in an airport. A photo of a gateway waiting area at an airport full of people.
}}
 \end{tabular} \\
  \bottomrule
\end{tabular}
\end{table*}
\begin{table*}
  \caption{Part 2 of coded locations. Shows the other four coded location levels, definitions, and examples applied under each level.}
  \label{table: locations2}
  \centering
    \begin{tabular}{p{4cm} p{4cm} p{3cm}}
    \toprule
    \textbf{Location} & \textbf{Location Examples} & \textbf{Image Example} \\                                                       \midrule
    Used in retail businesses    & Supermarkets, hair salons, and other businesses
     &  \begin{tabular}[c]{@{}p{0cm}@{}} 
\rotatebox{90}{\includegraphics[width=60pt, height=90pt]{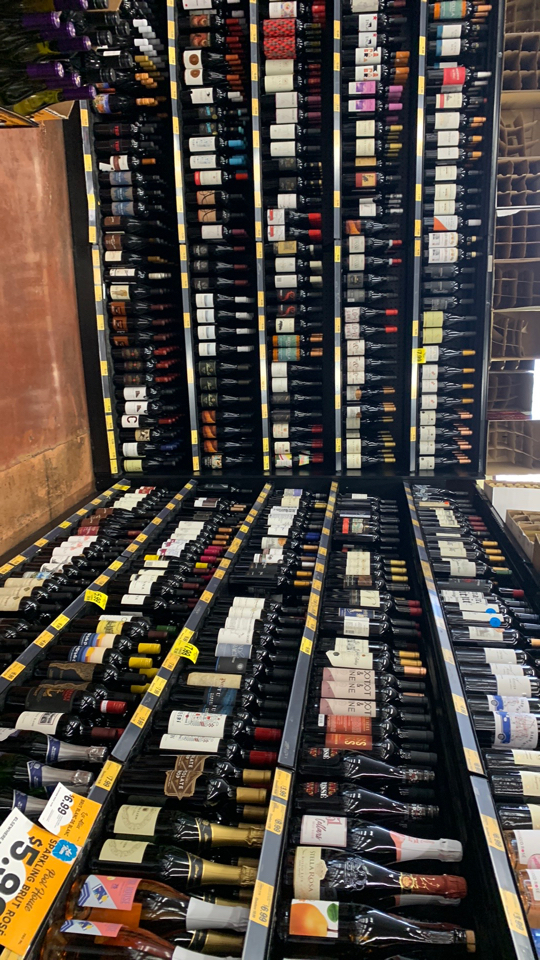}\Description{This photo represents an example of an entry coded for location “Used in retail businesses”. Participant was located in a supermarket, at the alcohol section. Shelves full of wine bottles at a supermarket
}}
 \end{tabular} \\
     Used at participant’s work    & Any spaces participants defined as their “work”, such as corporate or home offices
     &  \begin{tabular}[c]{@{}p{0cm}@{}} 
\rotatebox{0}{\includegraphics[width=60pt, height=90pt]{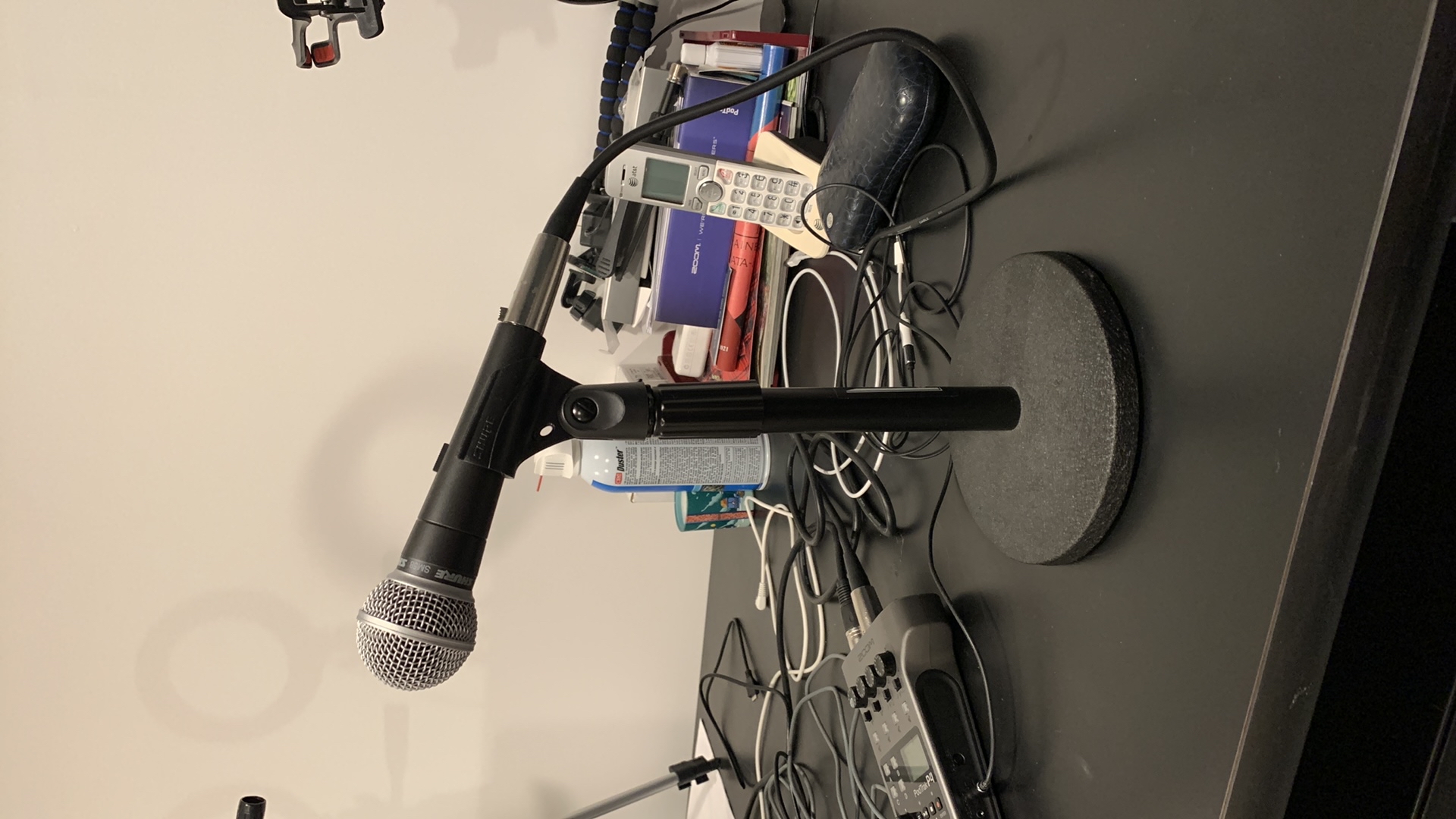}\Description{This photo represents an example of an entry coded for location “Used at participant’s work”. Participant was in their home studioin recording their podcast. A photo of a black and silve microphone on a stand with several electronic devices on the table
}}
 \end{tabular} \\
     Used in living spaces  & Apartments, senior homes, and houses
     &  \begin{tabular}[c]{@{}p{0cm}@{}} 
\rotatebox{0}{\includegraphics[width=60pt, height=90pt]{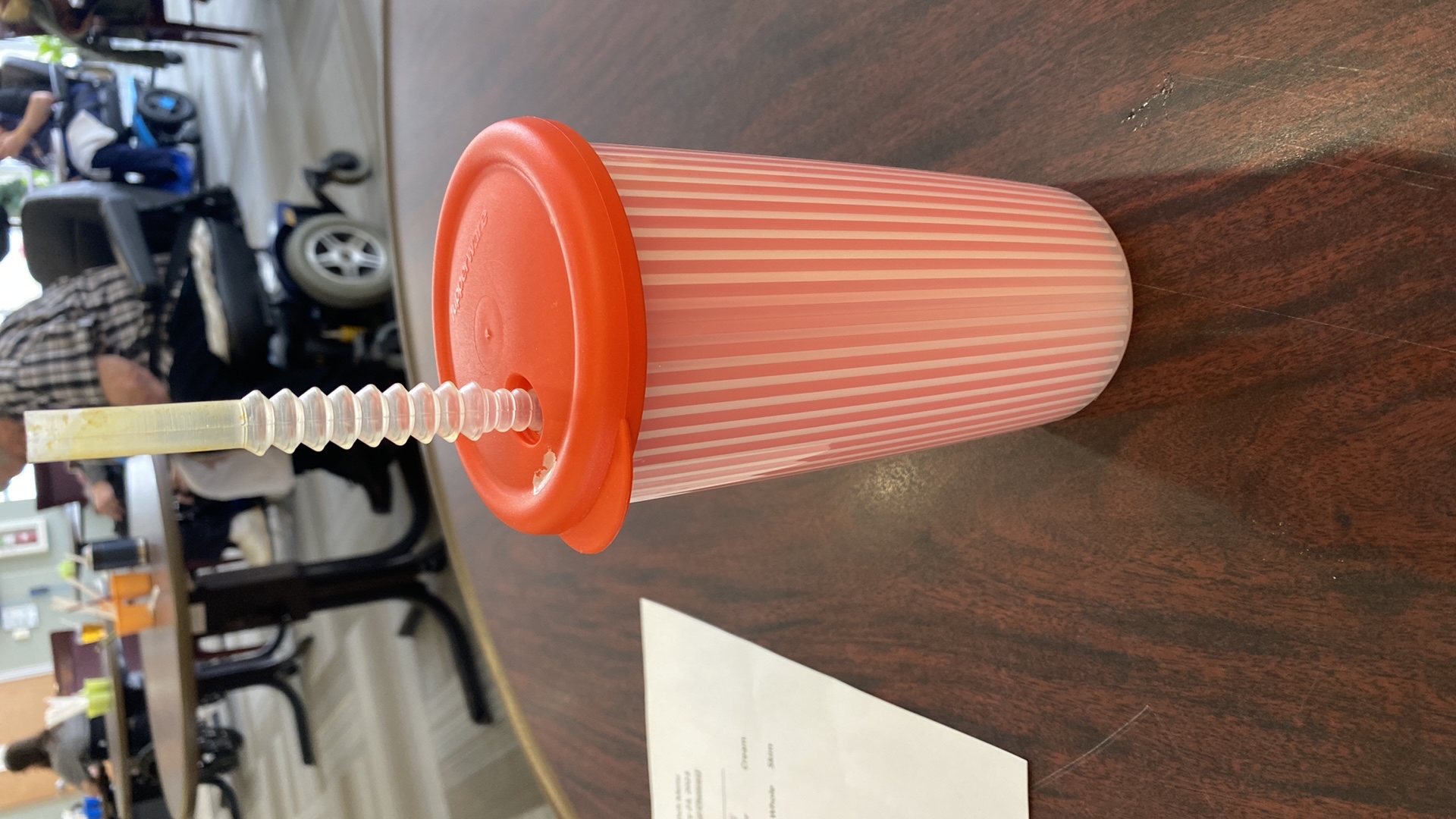}\Description{This photo represents an example of an entry coded for location “Used in living spaces or domiciles”. Participant was located at an assisted living facility. A photo of a stripe patterned drinking cup on a table. In the background there are people on wheel chairs.
}}
 \end{tabular} \\
     Used in unidentified locations    & Applied when neither the picture nor the participant’s diary responses could tell us where they were located
     &  \begin{tabular}[c]{@{}p{0cm}@{}} 
\rotatebox{0}{\includegraphics[width=60pt, height=90pt]{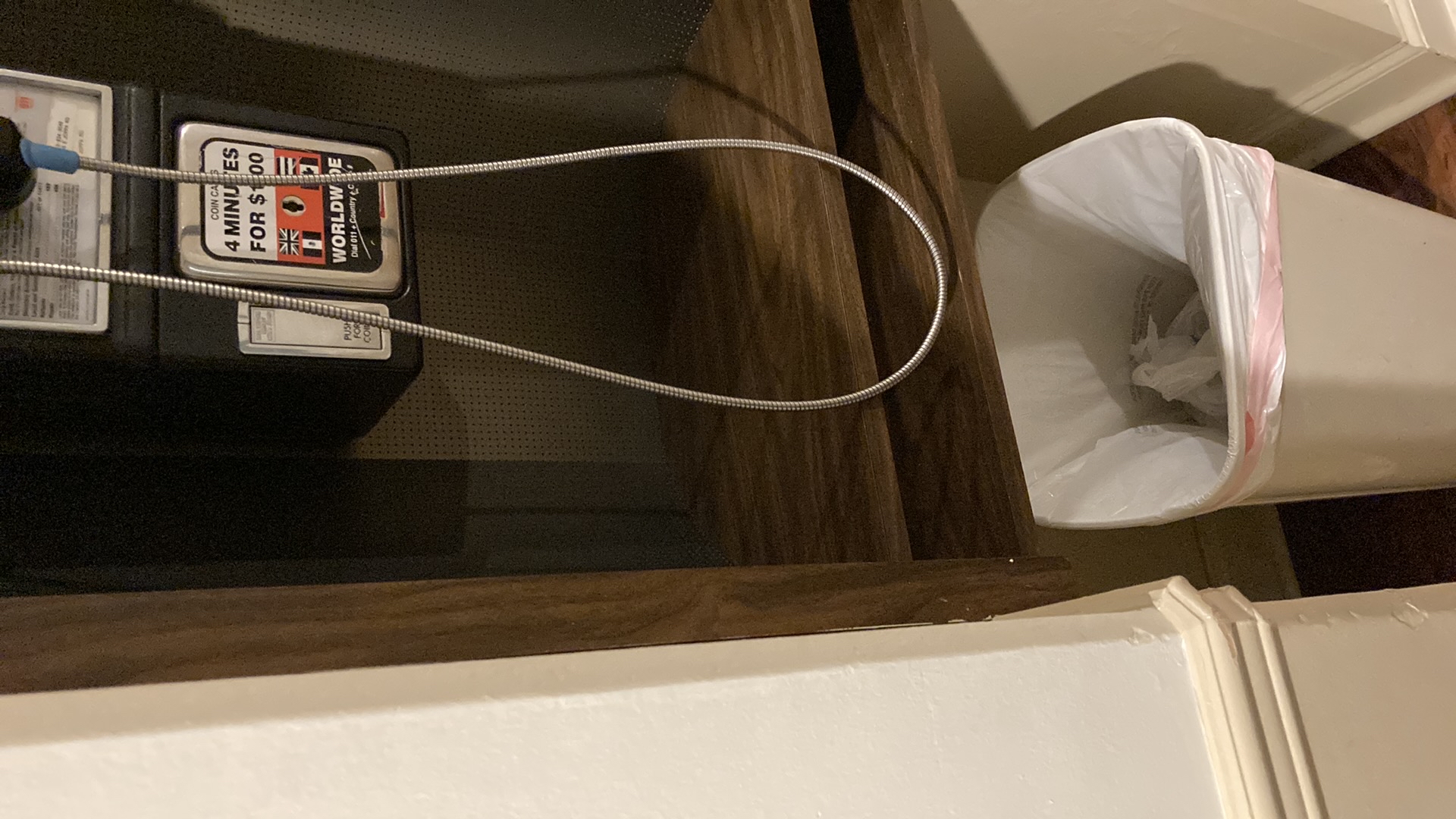}\Description{This photo represents an example of an entry coded for location “Used in unidentified areas”. Participant was located in a indoor space setting but in their dairy entry they did not mention where they were. A photo of a payphone and a trash can.
}}
 \end{tabular} \\
  \bottomrule
\end{tabular}
\end{table*}

To code follow-up interviews, audio recordings were transcribed by an automatic transcription service and corrected by one researcher. Then, two researchers coded four interviews individually. Finally, the researchers discussed disparities to add interview-derived codes to the codebook, and remaining interviews were split between researchers for coding.

These processes resulted in over 200 codes in total, which we grouped using multiple rounds of affinity diagramming. One researcher established three initial categories for code groupings: (1) Properties of the photos, (2) Content of the photos, and (3) Uses of the application. These were refined into subcategories (e.g., subcategories for objects captured as photo content such as living subjects, recreational objects and decor, etc.). Afterwards, we conducted thematic analysis using two rounds of affinity diagramming, one that examined grouped codes solely from the diary entries, and a second that incorporated interview codes. This generated our final themes (see Section \ref{Overview of Use Cases} and \ref{Choosing Between AI or Human Assistance}).

After coding and creating themes, \textbf{two researchers} also scored the accuracy of each entry’s scene description. We wanted to determine the appropriateness of the application descriptions given what information was available in submitted photos. This accuracy score allowed us to: (1) understand whether a scene description contained “usable” information, (2) understand how frequently the scene description application provided “reasonable” descriptions, and (3) find relationships between participants’ reactions and the accuracy of the descriptions. We defined accuracy as, “the extent to which the scene description matched a sighted human perception of the image,” adapted from user-based evaluations of image descriptions \cite{kreiss-contexual-metrics}. Two researchers coded two participants' diaries individually, then came together and split remaining entries, scoring each image-interpretation pair on a scale of 1 to 3, where 1 meant the interpretation did not match a human perception at all, 2 meant a partial match, and 3 meant it mostly matched. We share results about accuracy scores and discuss how participants’ definitions of satisfaction varied based on description accuracy (see Section \ref{Trust, Satisfaction, and Accuracy}).

\subsection{Quantitative Analysis} \label{Quantitative Analysis}

The goal of our quantitative analysis was to identify any significant effects on participants’ evaluative scores resulting from two distinct variables of use cases: user goal and location. We selected these two variables as these would be  most insightful towards understanding future use of AI-powered scene description applications and what kind of content it should excel or improve at describing. We conducted statistical tests with our quantitative models to find: (1) whether participants were scoring descriptions differently (satisfaction and trust) across identified user goals, and (2) whether the location where participants used the application had a significant effect on how they scored the descriptions (e.g., finding whether the application described scenes better at specific locations).

To represent user goals as a fixed effect in our model, we did another brief round of coding the diary entries so that each entry had just one user goal. We identified the main goal that participants reported in their entry based on their diary entry response, and from interview commentary.

We used linear mixed effects models (LMEM) fitted by Restricted Maximum Likely (REML) estimation to model all of our data. We conducted a post-hoc pairwise analysis to determine whether the satisfaction or trust participants felt after using the application was significantly different depending on their goal or location, with \textit{ParticipantID} as a random effect.

\textbf{\textit{Effects of User Goal and Accuracy.}}
 We explored the effect of participants’ goals on their satisfaction with and trust in the application. We fit one model with two factors, \textit{User Goal} (five levels: \textit{describeScene, learnApplication, identifySubject, identifyFeatures, noSpecificGoal}) and \textit{Accuracy}, and one measure, \textit{Satisfaction}. Each level in \textit{User Goal} was pulled from subcategories in our code groupings (see table \ref{table: goal-normalized} for examples and definitions). \textit{Satisfaction} was pulled from participants’ diary entries, while \textit{Accuracy} was the coded image-description score (see Section \ref{Qualitative Analysis}). Then we fit a second model with the same two factors, but this time with \textit{Trust} as a measure. \textit{Trust} was pulled from participants’ diary entries.

\textbf{\textit{Effects of Location of Use and Accuracy.}}
We analyzed the effect of the participants’ location while taking the photo on their satisfaction with and trust in the application. We fit one model with two factors, \textit{Location} (eight levels: \textit{usedInRecreationalOrLeisureAreas, usedInHealthcareSites, usedInEducationalOrSpiritualSites, usedInTravelServiceSites, usedInRetailBusinesses, usedAtParticipant’sWork, usedInLivingSpacesOrDomiciles, usedInUnidentifiedAreas}) and \textit{Accuracy}, and one measure, \textit{Satisfaction}. Each level in \textit{Location} was pulled from subcategories in our code groupings (see table \ref{table: locations}, and \ref{table: locations2} for examples and definitions).  Then we fit a second model with the same two factors, but this time with \textit{Trust} as a measure.

In our post-hoc analysis, since we made multiple comparisons between several estimates in our models, we performed p-value adjustments using the Tukey method for post-hoc analysis. In addition, since our data had independent sample variances between factors, all of our pairwise analysis t-tests were run with the Satterthwaite method to account for our complicated covariance. We share results from our statistical tests, and explain the effects that user goal and location had on participants' trust and satisfaction scores (see Section \ref{Trust, Satisfaction, and Accuracy}).

\section{Findings} \label{Findings}

In this section, we provide an overview of use cases identified in participants’ diary entries, describe participants’ preferences between human and AI assistance, and present an overview of satisfaction, trust, and accuracy scores across entries.

\subsection{Overview of Use Cases}\label{Overview of Use Cases}

We collected various data factors through diary entries that, together, comprised participants’ use cases for the application. We present a holistic overview of these factors: user goals, photo content, photo-taking locations, and photo quality.

\subsubsection{User Goals} \label{User-Goals}

\paragraph{\textbf{\textit{Overview of User Goals.}}} The user’s goal was an important component of identifying use cases because it revealed what visual information a participant wanted, as well as why they wanted it (e.g., they wanted a description of a light, specifically to know if it was on or off). In total, we identified 11 categories of user goals, most of which involved getting additional information about a subject participants were already somewhat knowledgeable about. For example, participants were aware they were holding a shirt, but they wanted to hear its color (more examples in table \ref{table: goal-normalized}). Participants sometimes mentioned multiple goals, which meant entries could fall under multiple categories.

Participants’ most common goal was to identify a specific subject (128 entries out of 316, 27\%). This occurred when they were looking for exact identification of a subject by name rather than a vague description (e.g., “television”’ instead of “large electronic device”) or were trying to identify a particular subject out of similar subjects (e.g., checking if they were holding their favorite yellow jacket instead of any jacket). Most participants already had some information about the subjects they were interested in when applying this goal. For example, if a participant was looking for a particular piece of clothing, they touched it to triangulate part of the information they needed along with requesting a description to determine what it was. This triangulation, which usually occurred before using the application, helped them utilize imperfect descriptions. Other goals included building understanding of the scenery (54 entries, 12\%), or reading text or numbers (38 entries, 10\%). In both of these cases, participants were usually asking about a subject they had some information on, but wanted additional specifics for, such as the scent label on a self-care product, or an aesthetic description of a mountain they knew was in the distance.

\paragraph{\textbf{\textit{Unique Goals.}}} Some participants had very unique visual information goals. For example, P11 took a picture of her pets to capture a sentimental moment. She explained: “My cat was on her back with my dog using her tummy as a pillow. I felt the way they were sleeping together was particularly enchanting and wanted it captured with a description.” P14 demonstrated another unique goal. She used the application to determine whether other people were around her. In one entry, she noted that she was thinking of entering a chapel, and she “wanted to find out if there was anyone in there so that I could go in by myself.” She determined that since the description had not mentioned people, the chapel must have been empty. Such goals highlighted the creative ways our participants used the application, and showed the descriptions could enhance their daily lives by supplying useful information.
\begin{figure}
\includegraphics[width=0.45\textwidth]{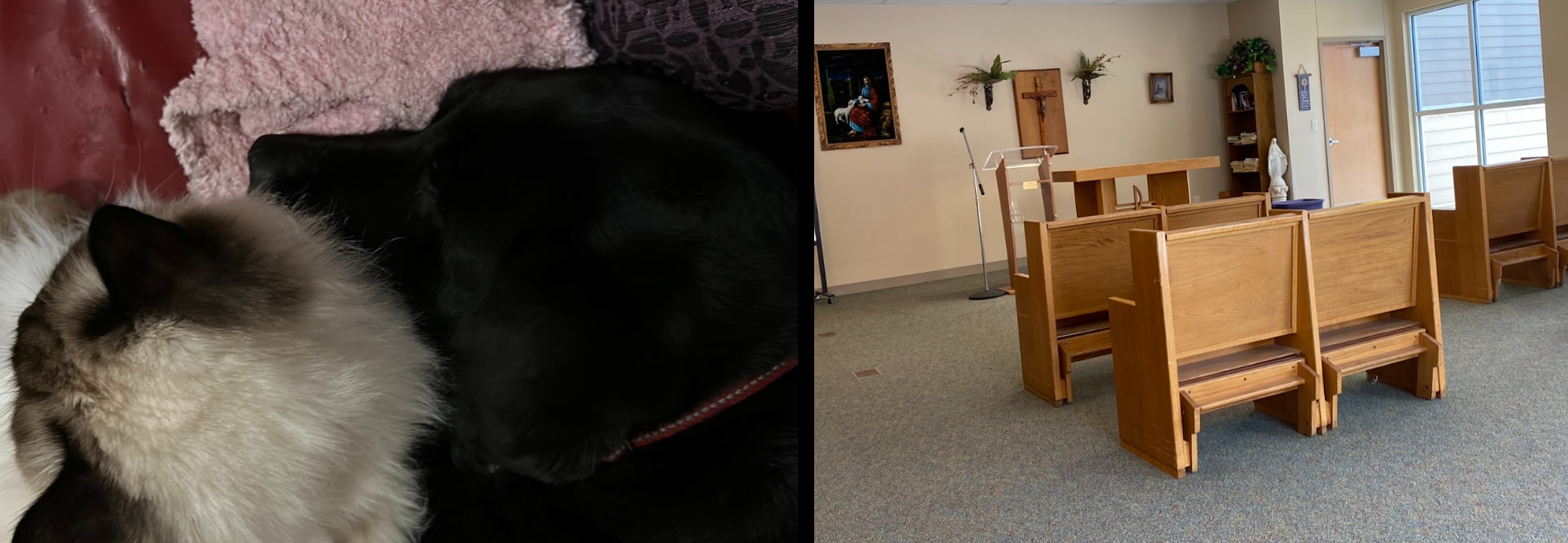}
\caption{\label{fig: uniquegoals}Two images exemplifying unique user goals. The image on the left, submitted by P11, shows a black dog's head laying on a white cat. P11 had wanted an interpretation to narrate the sentimental moment. The image on the right, submitted by P14, shows an empty chapel with wooden pews. P14 had wanted an interpretation to determine the privacy she would have in the space by checking for others' presence.
}
\Description{Two images exemplifying unique user goals. The image on the left, submitted by P11, shows a black dog's head laying on a white cat. P11 had wanted an interpretation to narrate the sentimental moment. The image on the right, submitted by P14, shows an empty chapel with wooden pews. P14 had wanted an interpretation to determine the privacy she would have in the space by checking for others' presence.}
\end{figure}

\begin{figure*}
\includegraphics[width=0.75\textwidth]{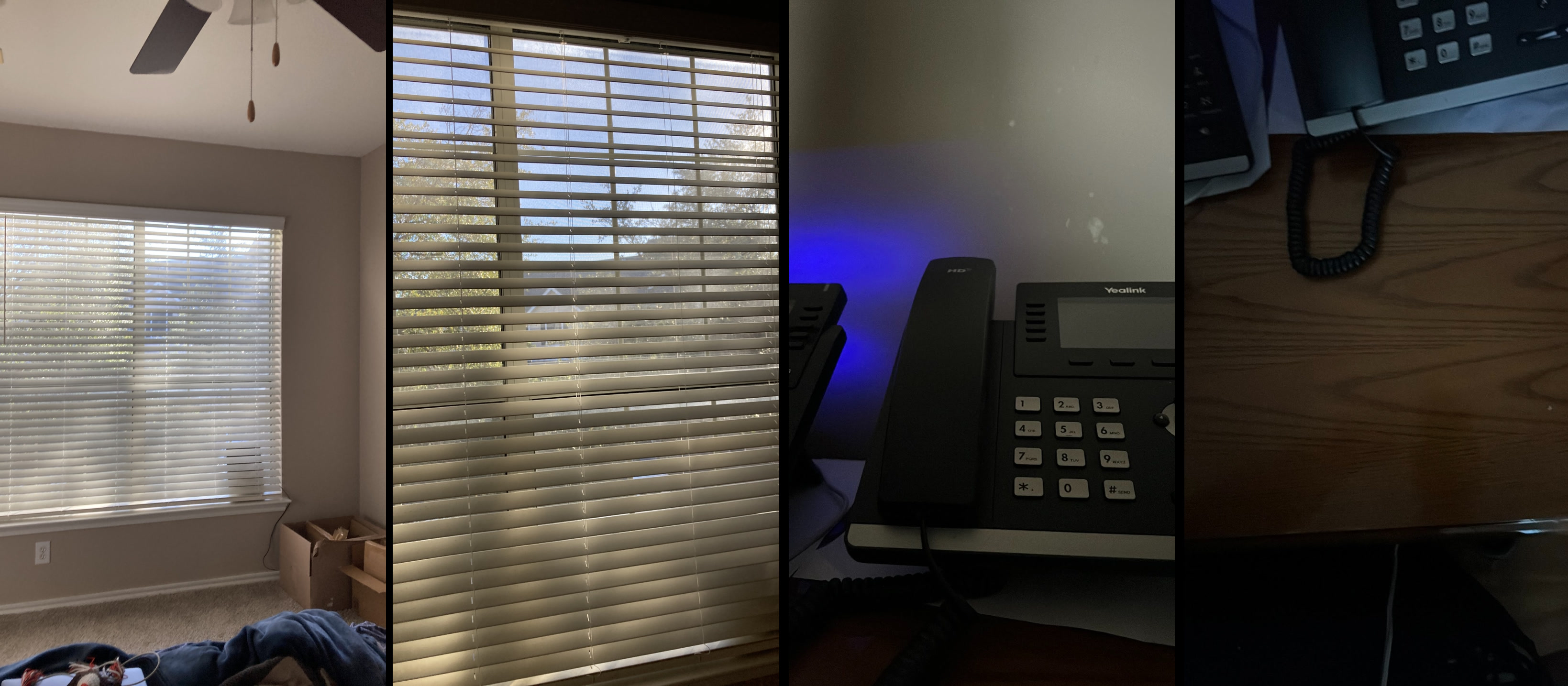}
\caption{\label{fig: connectedphotos}Examples of connected photos featuring the same subjects at varying distances; participants used the resulting interpretations to help them understand what was the best distance to obtain an accurate result. On the left, there are two photos of the same bedroom window submitted by P10. P10 wanted to learn how far away he needed to be from a photo subject for it to be fully captured and recognized by the application. On the right, there are two photos of a corded telephone submitted by P16, taken to “improve” her framing of the photo. 
}
\Description{Examples of connected photos featuring the same subjects at varying distances; participants used the resulting interpretations to help them understand what was the best distance to obtain an accurate result. On the left, there are two photos of the same bedroom window submitted by P10. P10 wanted to learn how far away he needed to be from a photo subject for it to be fully captured and recognized by the application. On the right, there are two photos of a corded telephone submitted by P16, taken to “improve” her framing of the photo.}
\end{figure*}

\paragraph{\textbf{\textit{Testing and Learning with the Application.}}} Some participants took photos not to receive new information from the application, but to test its results against their knowledge of familiar subjects. We called these “tester photos” and they were an unexpectedly common goal among our participants, motivating  38 photos (12\%). P3 provided an excellent example when he took a picture of his kitchen sink, explaining in his diary entry that he was “really disappointed that [the application] did not get any of it. Did not get the dishes did not get the silverware in the measuring cup or anything like that.” He was fully aware of all the items and wanted to know how well the application described them. P10 took several tester photos too, submitting multiple entries with photos of the same subject at different distances (see figure  \ref{fig: connectedphotos}). For instance, he took photos of a bedroom window at different distances, remarking that he was “just testing the app to see how it would make a recognition” and “get an idea of how far or close I should have the camera” for it to work properly. P16 took similar sets of tester photos, explaining, “One of the things I use artificial intelligence description apps for is to find the best position of the camera.” She further clarified that: “Sometimes I will take two or three pictures of the same thing from different distances and angles” to help her frame the photo (see figure \ref{fig: connectedphotos}).

\begin{table*}
  \caption{Definitions and examples of user goals identified in the diary entries. **The goal percentages are normalized as follows to avoid biasing the data toward participants who used the application more: $i = 16 \sum_{n=1} \left(\frac{E}{T} \times 100\right)$. Normalized percentage is equal to the sum of the number of each participant’s entries coded under a certain goal (E) divided by their total entries (T), times 100. Values are rounded up for simplified legibility. }
  \label{table: goal-normalized}
  \begin{tabular}{p{5.5cm}p{5cm}p{6cm}}
    \toprule
    \centering \textbf{User Goal (\% normalized**)} & \centering \textbf{Definition} & \textbf{Examples of Use}                                                        \\
    \midrule
    Identify Specific Object (27)   & Looking for the specific name of an object or trying to find a particular object based on its features.
     &  “I want to hear that it’s a television instead of a ‘large electronic device’ ”\newline
“I was looking for my favorite shirt”
 \\
 General Interest in Object Appearance (16)  & Participant did not mention a specific goal but was interested in receiving a description.
&  “I just want to know what it looked like". “Just playing around with the app.”
 \\
  Build Understanding of Scenery (12)   & Looking for a description of the environment or the weather.
&  “What do the mountains look like?” \newline
“Describe the building across the street.”
 \\
  Read Text or Numbers (10)  & Looking to read specific textual information within the photo (expected to be higher if our app performed OCR). &   “What does the label on this box say?” \newline
“Which scent is this soap?”
 \\
  Describe Digital Content (9)   & Looking to hear descriptions of the contents displayed on any kind of digital screen.
     &  “What was on the TV?” \newline
“Is my phone connected to my car’s Bluetooth? (looking for the symbol on the car’s screen)”
 \\
  Determine Visual Elements (8)   & Looking for visual features about objects like colors, shapes, sizes, patterns, contrast, or reflectiveness. &  “Am I looking at the red one?” “What size is the picture on this mug?”
 \\
  Test Application Against Current Knowledge (8)   & Testing the app to see how it worked, to see if it could answer “correctly”, or to see how descriptive it would be.
     &  “I wanted to see how close it would get.”
“Checking if it still recognized the telephone from far away.”
 \\
  Determine Subject Location or Presence (5)   & Looking for the location, orientation, or distance of a particular object in the camera frame, or checking if an object was present in the frame.
     &  “Where is the water bottle on my desk?” \newline
“How far am I from the stairs?”
 \\
  Determine Subject State or Amount (3)   & Asking about the state of an object or checking how much of an object was present in the frame.
     &  “Is the light on or off?” 
 \\
  Inform Camera Use (2) & Using the app to help them understand how the camera worked.
     &  “I wanted to know if I framed my potted plant correctly.” \newline
“Is the lighting here good enough for a nice photo?”
 \\
  Understand Person’s Appearance (1) & Looking for a description of their own or another person’s physical appearance.
     &  “I was curious to see how it would describe me.”
 \\
  \bottomrule
\end{tabular}
\end{table*}

\begin{figure*}
\includegraphics[width=0.95\textwidth]{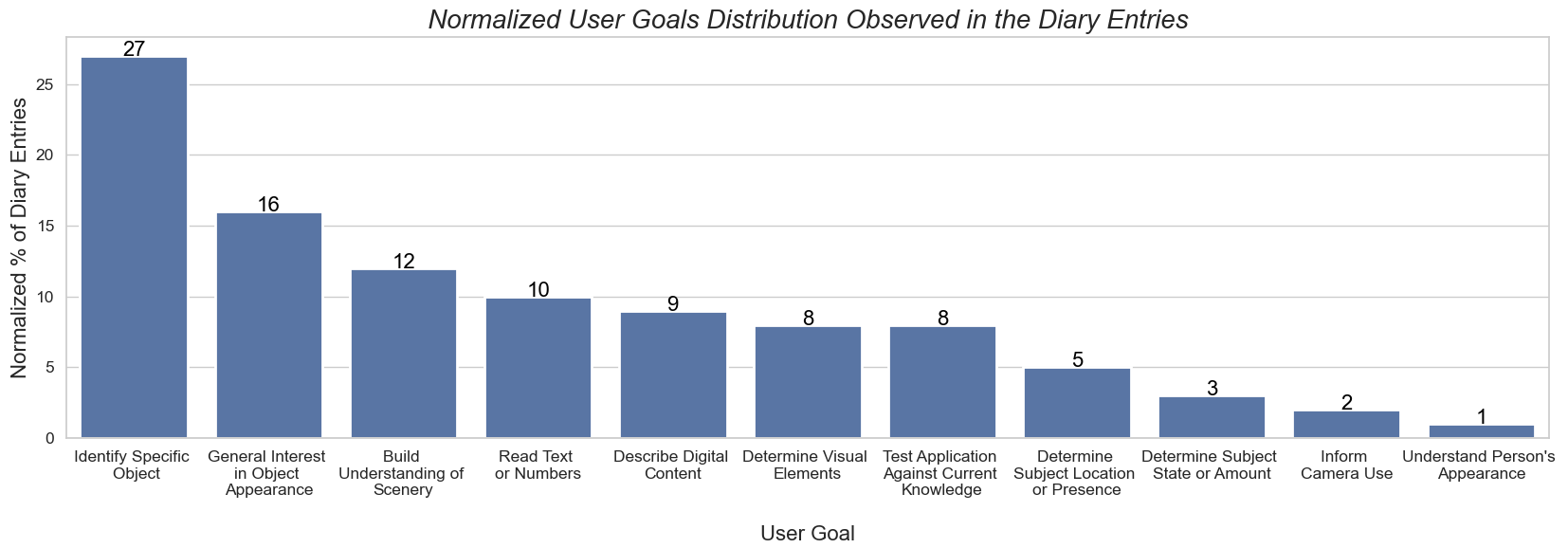}
\caption{\label{fig: normalizedgoals}Normalized frequency of user goals observed in the diary entries. Each value represents how frequently user goals happened (same as Table \ref{table: goal-normalized}). Values are rounded up for simplified legibility.
}
\Description{Normalized frequency of user goals observed in the diary entries. Each value represents how frequently user goals happened. Identify Specific Subject happened was the goal 27\% of the time, General Interest in Subject Appearance 16\%, Build Understanding of Scenery 12\%, Read Text or Numbers 10\%, Determine Visual Elements 8\%, Test Application Against Current Knowledge 8\%, Describe Digital Content 9\%, Determine Subject Location or Presence 5\%, Determine Subject State or Amount 3\%,  Inform Camera Use 2\%, and Understand person’s Appearance 1\%.}
\end{figure*}
\subsubsection{Content}

\paragraph{\textbf{\textit{Overview of Content.}}} The photo content provided information about the kinds of subjects our participants captured to address their use cases (e.g., to know if a room was messy, they photographed the floor and any objects on it). In total, we established 18 content categories based on objects position and function (Stationary Subjects, Immobile Objects, Objects within Reach, Objects out of Reach, Symbolic Objects, Digital Content or Screens, Landscapes, Cleaning and Medical Supplies, Text and Labels, Recreational Objects, Decorative Objects, Living Subjects, Clothing and Self-care, Kitchen Supplies, Vehicles, Keys, Building Architecture, and Electronic Devices). The content participants captured showed a focus on stationary subjects, often ones they were capable of walking up to or holding easily. We defined stationary subjects as subjects that were immobile, or were not likely to move after the photo was taken. For example, a parked car was considered “stationary” if it was turned off and waiting in a parking lot, but a car driving along a road was considered “moving.” The great majority of entries, 299 (95\%), included at least one stationary subject, whereas only 23 entries (7\%) contained at least one moving subject. We coded all subjects that were readily identifiable in each photo, meaning photos could have multiple content codes that counted as stationary or moving.
\begin{figure}
\includegraphics[width=0.2\textwidth]{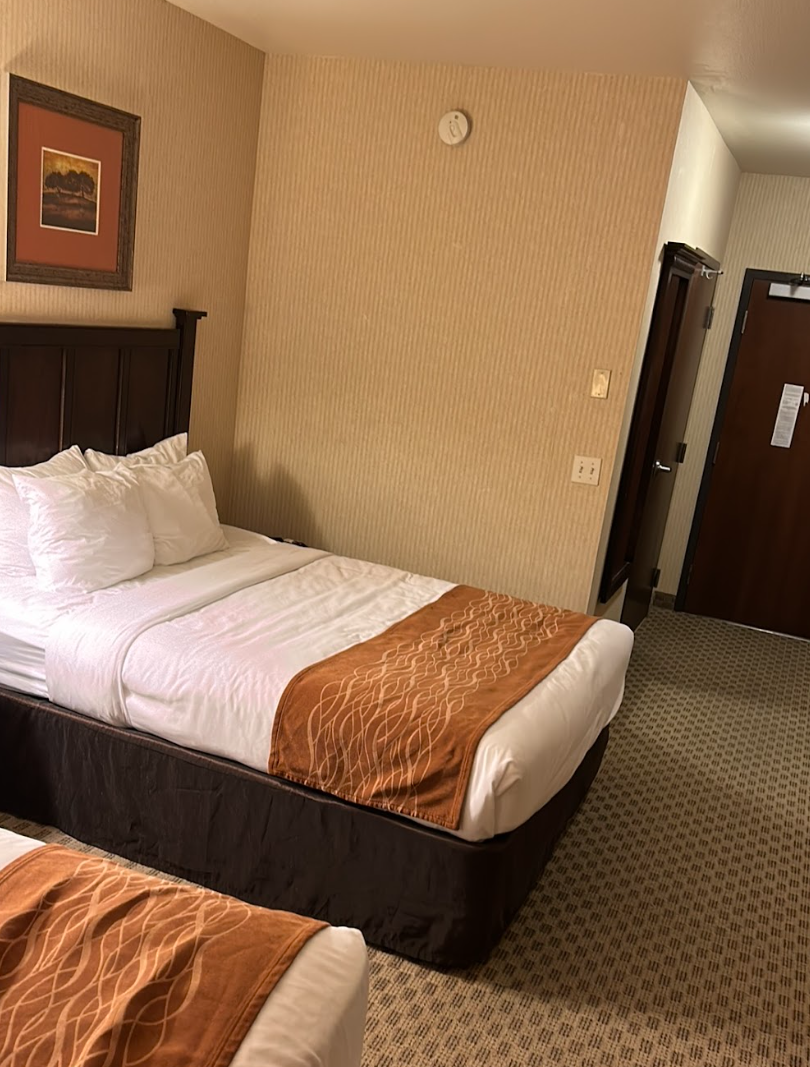}
\caption{\label{fig: avoidtouch}A photo of an empty hotel bed, submitted by P5 to verify that his roommate was not sleeping in the bed. In this case, P5 took this photo because he did not want to inspect with his hands whether his roommate was present in the bed. 
}
\Description{A photo of an empty hotel bed, submitted by P5 to verify that his roommate was not sleeping in the bed. In this case, P5 took this photo because he did not want to inspect with his hands whether his roommate was present in the bed.
}
\end{figure}
The top three most common content categories were: decor or decorative objects (131 entries, 41\%), living subjects (105 entries, 33\%), and building architecture or built-in structures (96 entries, 30\%). Decor or decorative objects contained wall decorations, posters, and general home decor like stylized clocks. Living subjects covered plants (which were sometimes also decor, but not always, e.g., trees lining the side of a road, home gardens of flowers, vegetables, or herbs) and creatures like humans or animals (e.g., cats and dogs). Building architecture included things like stairs, ramps, or street lights. Of the wide variety of content captured here, only living subjects contained moving subjects. From this, we conclude that our participants were largely focused on stationary subjects unless taking pictures of a pet, themselves, or another person.

\paragraph{\textbf{\textit{Inspecting Inconvenient (to touch) Objects.}}} One of the most interesting and commonly-captured characteristics in participants’ diary entries were objects they thought were inconvenient to inspect directly with their hands. This could mean the object was too far to touch, was possibly dirty or disgusting, or even considered dangerous to touch. For example, P15 submitted an image at her local grocery store of a row of wine bottles. She wanted information about the wine, but also wanted to avoid knocking it over, so she took a picture instead of touching it. Other cases where participants did not want to touch objects involved not wanting to touch another person’s belongings, or making sure they were not accidentally touching other people. This was the case of P5, who took a picture of a hotel bed to determine “whether my roommate was there or not there cause I did not wanna touch the bed” in case his roommate was sleeping there already (see figure \ref{fig: avoidtouch}).

We discussed this topic with some participants in their follow-up interviews, who gave examples of when they had run into issues exploring things by touch outside our study. P1 mentioned one of her entries had reminded her of a memorable experience:

\newenvironment{myquote}
{\list{}{\leftmargin=0.15in\rightmargin=0.15in}\item[]}
{\endlist}

\begin{myquote}
I had been hanging out with my siblings and with my niece, and I had accidentally put my hand in a dirty diaper that had not been rolled up and thrown away. And it was disgusting. And I had touched it because I thought it was one of the toys. And so like getting that aspect of like a clear identification [with this photo entry] of what the object is like, this is a stuffed animal, and it is on the couch. For me, that made me feel a lot better, because I wasn't about to touch a dirty diaper. (P1, female, age 24)
\end{myquote}

There were many scenarios where our participants solely relied on scene description to take action. In some instances, they would avoid using touch due to social norms around their peers, or to avoid doing something inappropriate. In other cases, they would avoid using their hands due to personal hygiene or past bad experiences. This highlighted the interplay among societal expectations, individual preferences, and past encounters, all of which contributed to the different reasons participants engaged with the scene description application.

\subsubsection{Locations}

\paragraph{\textbf{\textit{Overview of Locations.}}} The photo location indicated the kinds of environments wherein our participants spent time regularly and also needed visual information, and thus, where they found use for the application. For example, they needed visual information while at the grocery store, often regarding objects they were considering buying. In total, we established 8 location categories (see table \ref{table: locations} and \ref{table: locations2}).

\begin{figure*}
\includegraphics[width=0.95\textwidth]{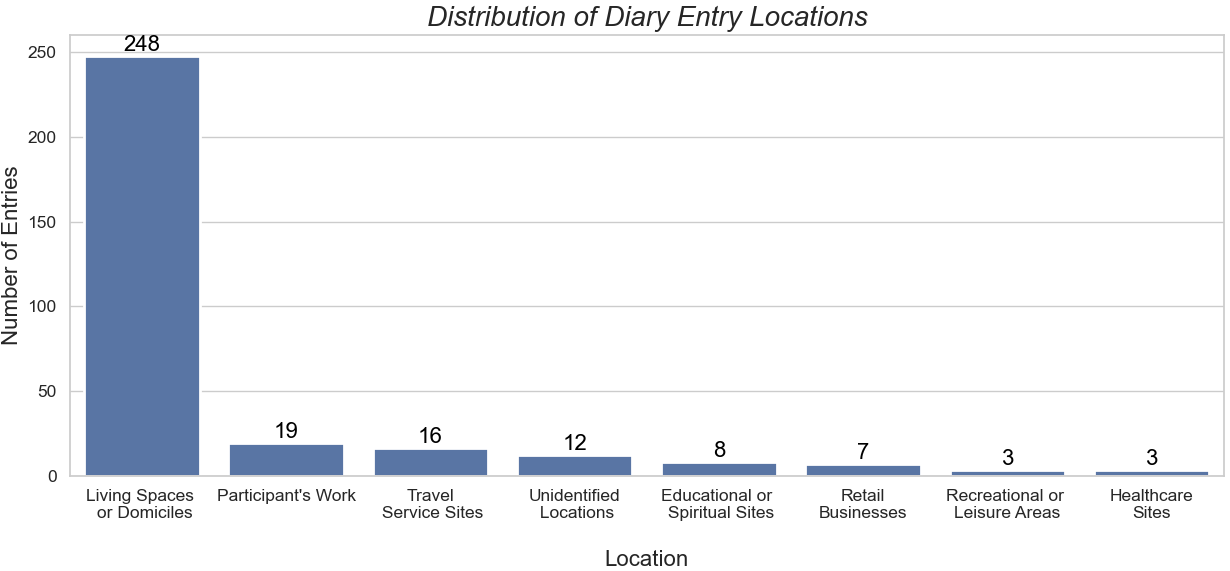}
\caption{\label{fig: distributionlocations} The distribution of locations identified in the diary entries.
}
\Description{The distribution of the locations identified in the diary entries. Living spaces is the most frequent category with 248 out of 316 diary entries having this code, and Healthcare sites is the least frequent category with 3 out of 316. All the other categories are: Educational sites 8, travel service sites 16, unidentified locations 12, recreational areas 3, retail businesses 7, and participants’ work 19.
}
\end{figure*}

Most participants’ photos were taken in familiar, indoor locations, the vast majority of which were living spaces like the participants’ homes. Conversely, this suggests that participants did not use the application in unfamiliar environments. By far, the most common location was living spaces (248 entries, 78\%), which included participants’ homes as well as ones they visited (see figure \ref{fig: distributionlocations}). The next two most common locations were at participants' work and travel sites (19 and 16 entries, 6\% and 5\% respectively), which included participants' home offices, public workplaces, personal vehicles, as well as bus stops, train stations, or airports. Many of these spaces are public, indoor buildings that we would expect our participants to be very familiar with from their daily commutes (e.g., local train stations, their own workplaces). As such, all of these locations comprised areas we would expect participants to be familiar with, confirming the trend we noted above.

\subsubsection{Photo Quality}

\begin{figure}
\includegraphics[width=200pt]{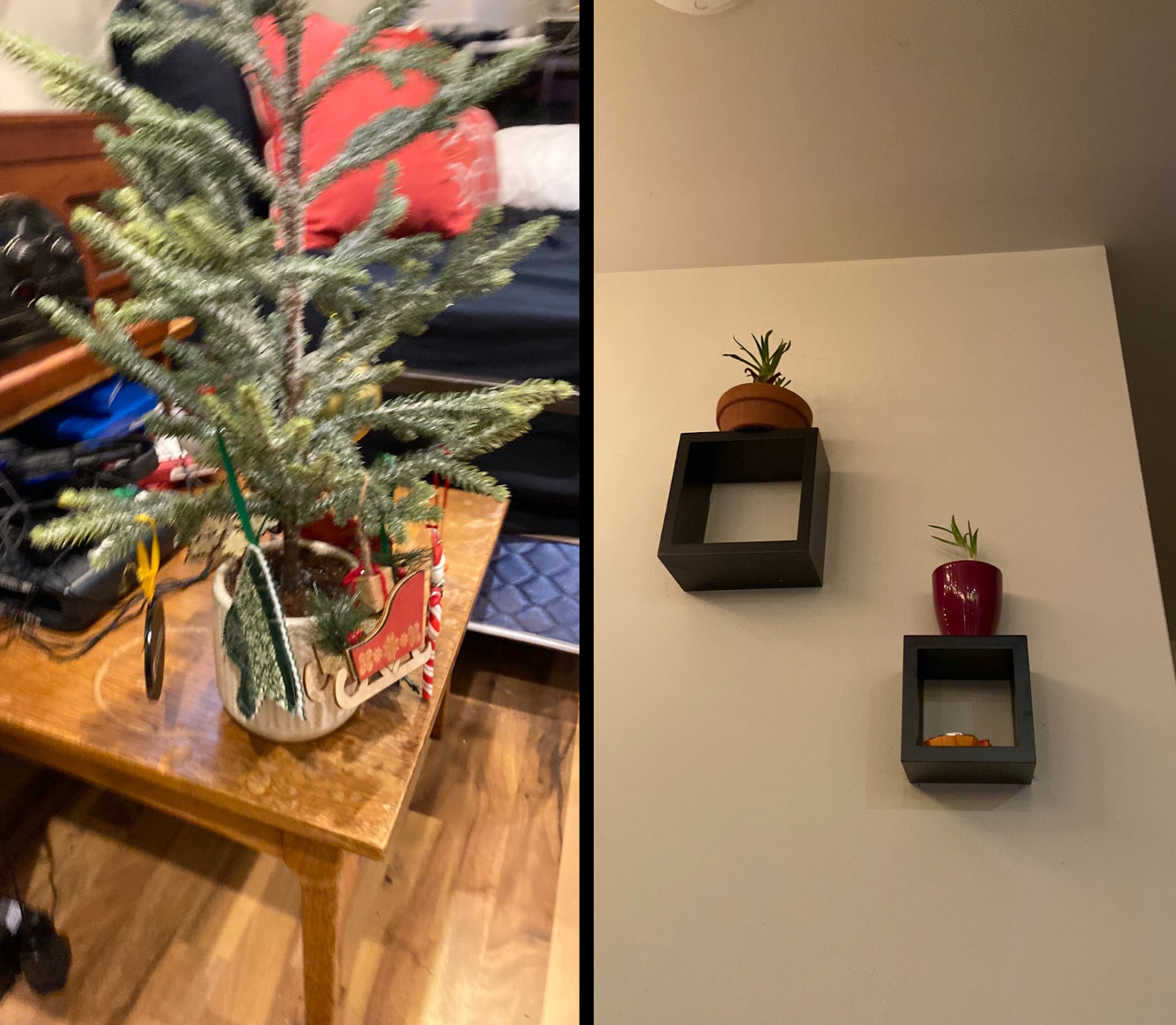}
\caption{\label{fig: mixedconditions}Two images exemplifying mixed photo conditions. On the left, a Christmas tree stands in a bright room with blurry branches. On the right, two plants rest  on decorative shelves, which cast shadows on the wall behind them. Both submitted by P4.
}
\Description{Two images exemplifying mixed photo conditions. On the left, there is a Christmas tree in a bright room, with very blurry branches. This photo shows low clarity with good framing and good lighting conditions. On the right, there are two plants resting on decorative shelves, which cast shadows on the wall behind them. This photo shows high clarity with good framing and mediocre lighting conditions. Both submitted by P4.
}
\end{figure}

\paragraph{\textbf{\textit{Overview of Photo Quality.}}} The photo quality provided us information about how participants took photos to address their use cases, which often differed from the way a sighted individual would take a photo (e.g., to get a description of a Christmas tree ornament, a participant would take the photo extremely close-up to the ornament). This information also provided an idea of what level of photo quality our participants were submitting, which impacted the descriptions generated by the application’s CV model. Most photos submitted had poor lighting, but were not blurry and had unoccluded subjects (they were well-framed). Out of 316 diary entries, 290 (92\%) were clear, 162 (52\%) had fully visible, unoccluded subjects, and 201 (64\%) had poor lighting conditions, which ranged from complete darkness to uneven lighting that created shadows on photo content. A common example of poor lighting was subjects being side-lit or back-lit, making them into silhouettes or obscuring part of their features. When looking at a single photo, the quality of these conditions were often mixed. For example, a photo could have been clear but was taken in poor lighting, like a clear picture of wall decorations where lighting casts shadows across the wall. Similarly, a photo could have been well-lit, but lacked clarity, like a close-up image of a Christmas tree in a well-lit room, with the tree coming out blurry due to the participant's action (see figure \ref{fig: mixedconditions}).

\subsection{Choosing Between AI or Human Assistance}\label{Choosing Between AI or Human Assistance}
During the follow-up interviews, we learned how participants chose whether to use the study application or other information resources, primarily human assistants. We highlight these observations below.

\textbf{\textit{Reasons for Using AI Tools Instead of Human Assistants.}} Participants gave various reasons for choosing AI tools over human assistants. Among the most common was that when human assistants were not around, they preferred to use AI instead of video calling someone to \textbf{address simple problems}. For example, P1 submitted a photo of a pair of sunglasses when she “was not sure if this was my husband’s sunglasses or my old pair of reading glasses from 4 years ago.” She was happy to get an answer without having to call her husband. P9 supported this in her interview, explaining that “I like being independent […] I have a lot of family and I have a lot of friends. But […] I'm not going to go out of my way and ask them when I have a piece of technology that can describe it to me.”

Interestingly, some participants expressed that since they believed AI would be unbiased, they could use it to \textbf{solve disputes about visual information}. We saw this with P5, who submitted a photo taken through an airplane window, with the goal of “find[ing] out if we were flying near the strip or close to the airport when taxiing.” He elaborated in his interview that he had been flying with a BLV friend, and they were arguing about the view outside. Since they believed AI would give the unbiased truth, they used it to settle the issue without having to ask another human third-party.

\textbf{\textit{Reasons for Using Human Assistants Instead of AI Tools.}} Along with such reasons for using AI, participants also shared reasons they preferred human assistants (aside from obvious ones such as higher accuracy or the ability to reliably communicate their information preferences). In particular, since \textbf{AI could not give thoughtful opinions} on content, some participants would not use the application to verify images they planned to post on public platforms like social media. P12 gave her thoughts on this in our interview:

\begin{myquote}
So it is usually a combination of having that image, but then also verifying with somebody else, if it is something that is really important, or something that I am going to, like, share on social media, or you know, to the public or something. I want to make sure that it is clear. So I would still verify after using Seeing AI with somebody else like with Aira or with another person. (P12, female, age 32)
\end{myquote}

These examples show how participants considered pros and cons of AI and human assistants, deciding where each option’s strengths most minimized potential errors or strains on human relationships before moving forward.

\subsection{Trust, Satisfaction, and Accuracy}\label{Trust, Satisfaction, and Accuracy}

\subsubsection{Overview of Trust, Satisfaction, and Accuracy}

We present a descriptive overview of satisfaction, trust, and accuracy scores across diary entries. We describe the application’s performance for each user goal and location in terms of trustworthiness, satisfaction and accuracy. We also present participants’ definitions of trust and satisfaction, and the differences (or similarities) between these in the context of the application.

\begin{figure*}
\includegraphics[width=0.9\textwidth]{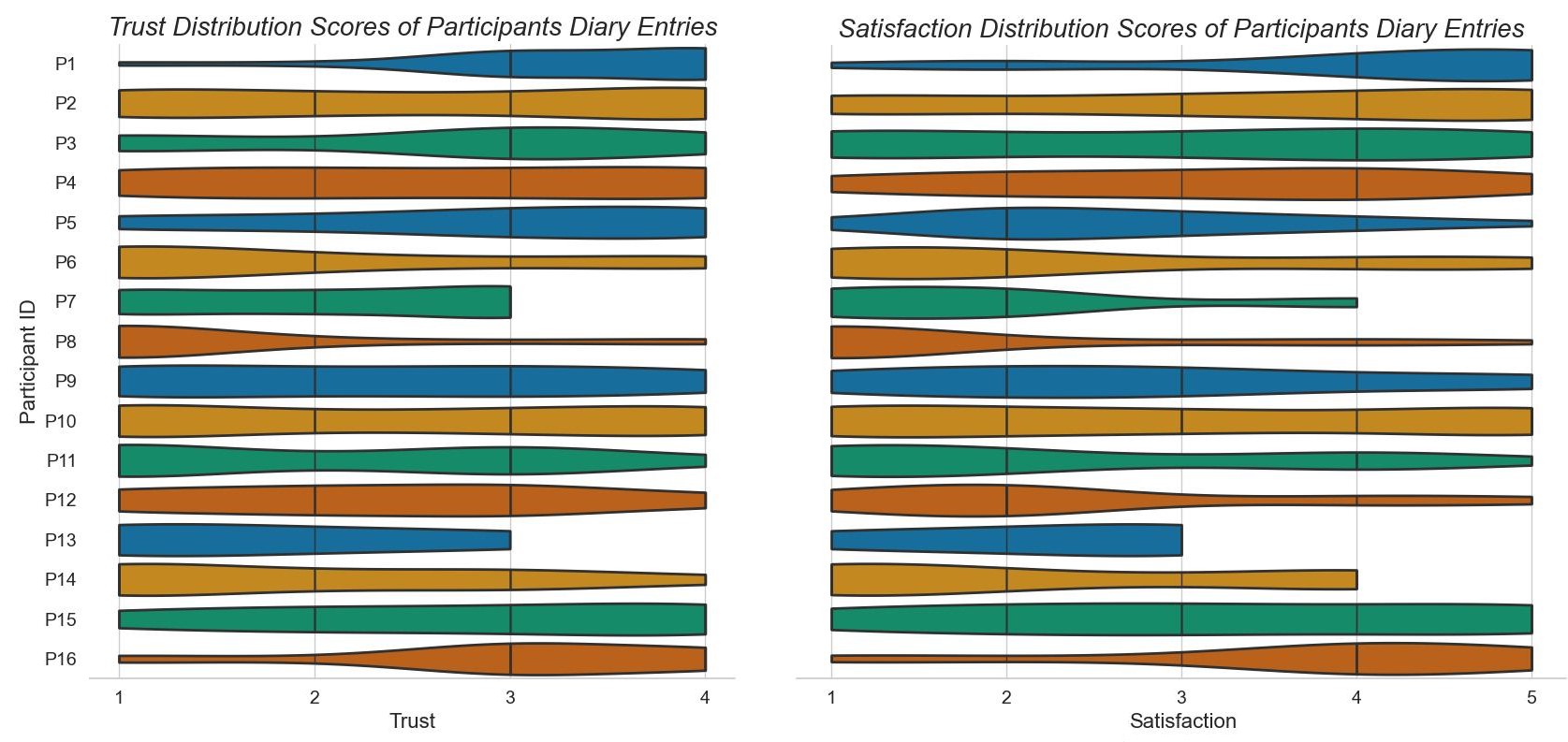}
\caption{\label{fig: violinplots}The distribution of participants' trust and satisfaction scores across their diary entries.
}
\Description{Violin plots displaying participants' satisfaction and trust scores distribution across their diary entries. P1 and P16 score distributions for satisfaction and trust are concentrated on the higher end of the scale (e.g., more satisfaction and trust). P2, P3, P4, P9, P10 and P15 scores distributions for satisfaction, and trust are all relatively evenly distributed; P2, P3, and P15 having slightly more diary entries with higher values. P6, P7, P8, P11, P13, and P14 scores distribution for satisfaction and trust are concentrated on the lower end of the scale. P5, and  P11 score distributions for satisfaction and trust are mixed; satisfaction scores are concentrated on the lower end of the scale, while trust scores are concentrated on the higher end of the scale.
}
\end{figure*}

\begin{figure}
\includegraphics[width=0.45\textwidth]{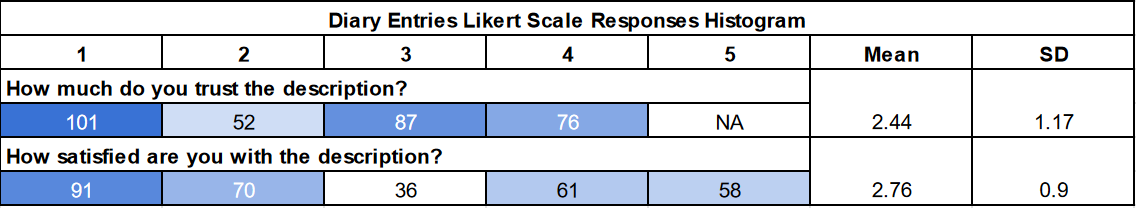}
\caption{\label{fig: histogramtrustsatisfaction} Diary Entries Likert scale scores for Trust and Satisfaction responses.
}
\Description{Diary entries Likert scale Trust and Satisfaction responses. The mean trust for all diary entries was 2.44 and the standard deviation was 1.17. 101 out of 316 Diary entries fell into the bucket of score 1, meaning participants did not trust those descriptions at all. 52 Diary entries fell into the bucket of score 2, meaning participants trusted those descriptions a little. 87 Diary entries fell into the bucket of score 3, meaning that participants trusted those descriptions somewhat. 76 Diary entries fell into the bucket of score 4, meaning that participants trusted those descriptions to a great extent.

The mean satisfaction for all diary entries was 2.76 and the standard deviation was 0.9. 91 out of 316 Diary entries fell into the bucket of score 1, meaning participants were very dissatisfied with  those descriptions. 70 Diary entries fell into the bucket of score 2, meaning participants were somewhat dissatisfied with those descriptions. 36 Diary entries fell into the bucket of score 3, meaning that participants were not satisfied nor dissatisfied with those descriptions. 61 Diary entries fell into the bucket of score 4, meaning that participants were somewhat satisfied with those descriptions.  58 Diary entries fell into the bucket of score 5, meaning that participants were very satisfied with those descriptions.
}
\end{figure}

\begin{figure}
\includegraphics[width=0.45\textwidth]{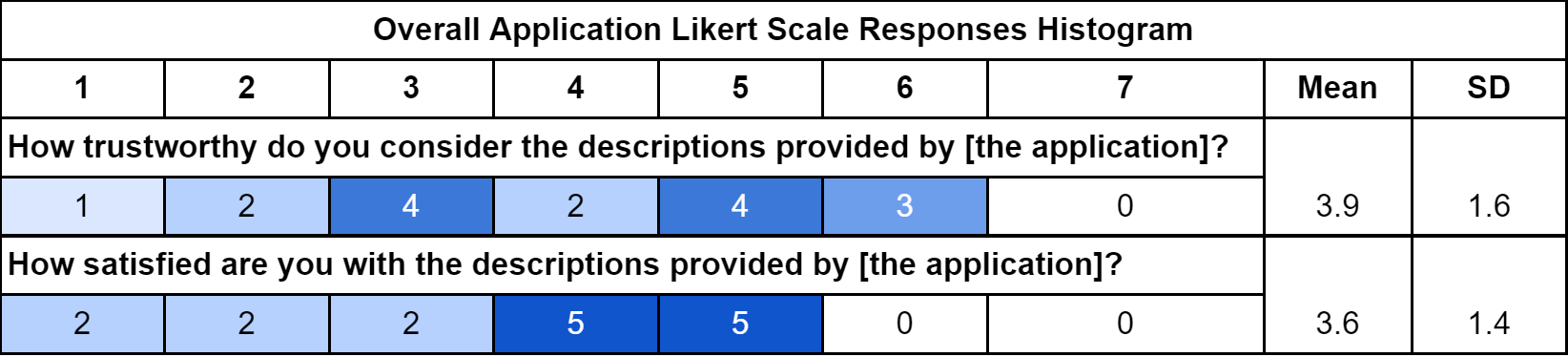}
\caption{\label{fig: histogramoveralltrustsatisfaction} Overall Application Likert scale scores for Trust and Satisfaction responses.
}
\Description{Overall Application Likert scale Trust and Satisfaction responses. The mean trust for the application was 3.9 and the standard deviation was 1.6. One participant thought the application was very untrustworthy. Two participants thought the application was somewhat untrustworthy. Four participants thought that the application was a little untrustworthy. Two participants thought the application was not trustworthy nor untrustworthy. Four participants thought the application was a little trustworthy. Three participants thought the application was somewhat trustworthy. No participants thought the application was very trustworthy.

The mean satisfaction for the application was 3.6 and the standard deviation was 1.4. Two participants thought the application was very dissatisfying. Two participants thought the application was somewhat dissatisfying. Two participants thought that the application was a little dissatisfying. Two participants thought the application was not satisfying nor dissatisfying. Five participants thought the application was a little satisfying . Five participants thought the application was somewhat satisfying. No participants thought the application was very satisfying.

}
\end{figure}

\paragraph{\textbf{\textit{Overview of Trust and Satisfaction Ratings.}}} Participants reported relatively low scores for trust and satisfaction, indicating that current commercially available CV models still have a long way to go to deliver satisfying and trustworthy descriptions for BLV users. On average, diary entries descriptions received a score of 2.43 on a 4-point scale (SD=1.16) for trust and 2.76 on a 5-point scale (SD=1.49) for satisfaction (see figure \ref{fig: histogramtrustsatisfaction}). In addition to scores reported in the diary entries, we also collected ratings on a 7-point Likert scale during participants’ interviews for how trustworthy and satisfying the application was overall. On average, it received a score of 3.9 for trustworthiness (SD=1.6) and 3.6 for satisfaction (SD=1.4). Like the diary entry scores, these remain in the average to below average range (see figure \ref{fig: histogramoveralltrustsatisfaction}). We unpack some of the reasoning and exceptions to these scores given by participants below.

\begin{figure}
\includegraphics[width=0.45\textwidth]{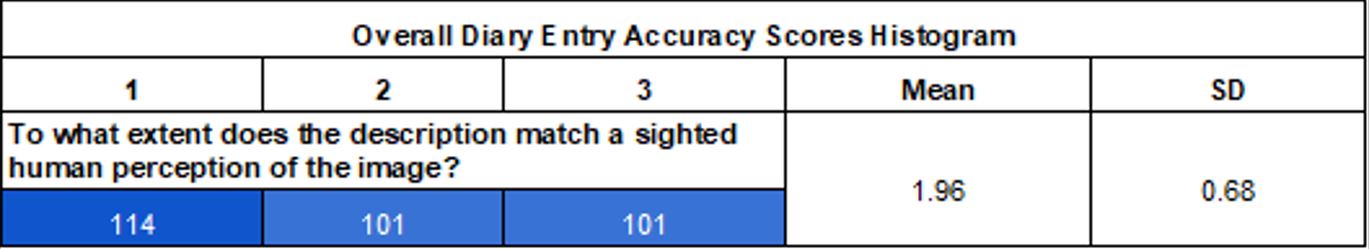}
\caption{\label{fig: histogramaccuracy} Distribution of accuracy scores for all the diary entries.
}
\Description{Accuracy score distribution across the 316  diary entries that received a description in our dataset. 22 entries were removed because they didn’t receive a description or had errors. The mean accuracy for all diary entries was 1.96 and the standard deviation was 0.68 . 114 out of 316 Diary entries fell into the bucket of score 1, meaning those descriptions did not match a human perception of the image. 101 Diary entries fell into the bucket of score 2, meaning those descriptions somewhat matched a human perception of the image. 101 Diary entries fell into the bucket of score 3, meaning those descriptions closely matched a human perception of the image.
}
\end{figure}

\begin{figure}
\includegraphics[width=0.45\textwidth]{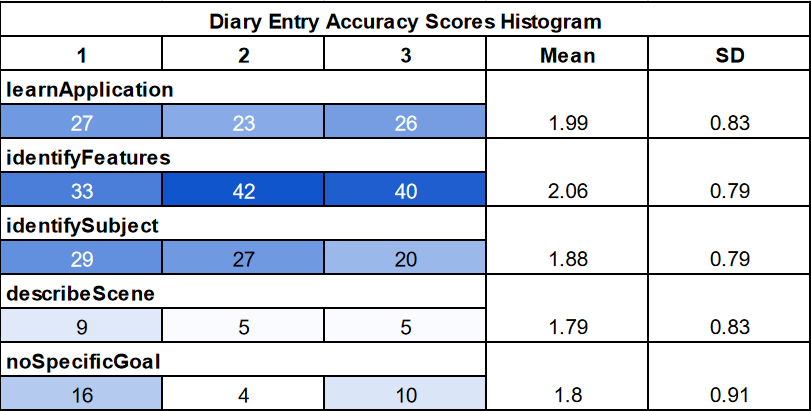}
\caption{\label{fig: histogramaccuracyusergoal} The distribution of accuracy scores for each user goal.
}
\Description{Accuracy score distribution of each user goal across the 316 diary entries analyzed in our dataset. The mean accuracy for the 76 entries coded with user goal Learn Application was 1.99 and the standard deviation was 0.83. 27 out of 79 Diary entries fell into the bucket of score 1. 23 Diary entries fell into the bucket of score 2. 26 Diary entries fell into the bucket of score 3.

The mean accuracy for the 115 entries coded with user goal Identify Features was 2.06 and the standard deviation was 0.79. 33 out of 115 Diary entries fell into the bucket of score 1. 42 Diary entries fell into the bucket of score 2. 40 Diary entries fell into the bucket of score 3.

The mean accuracy for the 76 entries coded with user goal Identify Subject was 1.88 and the standard deviation was 0.79. 29 out of 76 Diary entries fell into the bucket of score 1. 27 Diary entries fell into the bucket of score 2. 20 Diary entries fell into the bucket of score 3.

The mean accuracy for the 19 entries coded with user goal Describe Scene was 1.79 and the standard deviation was 0.83. 9 out of 19. Diary entries fell into the bucket of score 1. 5 Diary entries fell into the bucket of score 2. 5 Diary entries fell into the bucket of score 3.

The mean accuracy for the 30 entries coded with user goal No Specific Goal was 1.8 and the standard deviation was 0.91. 16 out of 30 Diary entries fell into the bucket of score 1. 4 Diary entries fell into the bucket of score 2. 10 Diary entries fell into the bucket of score 3.
}
\end{figure}

\textbf{\textit{Overview of Accuracy.}} The application’s accuracy scores were drawn from researchers’ coding of photo-description pairs (see Section \ref{Qualitative Analysis}). On average, accuracy was 1.95 on a 3-point scale (SD=0.82, see figure \ref{fig: histogramaccuracy}). This would suggest the application did not perform particularly poorly or well. We can also look at the accuracy per user goal, using the categories \textit{describeScene, learnApplication, identifySubject, identifyFeatures,} and \textit{noSpecificGoal} (see figure \ref{fig: histogramaccuracyusergoal}). These are the higher-level groupings of the goal categories from \ref{User-Goals}, developed for our quantitative analysis (see \ref{Quantitative Analysis}). We coded the Diary entries again to identify the one goal which reflected the main goal of the participant. These provided a better look at the distribution of accuracy scores, since entries often contained multiple goals from the full range in \ref{User-Goals} which would conflate the distribution. Per user goal, we see the application performed best on \textit{identifyFeatures} (2.06, SD=0.79), though it still did not reach a particularly high average. Similarly, the least accurate goals, \textit{noSpecificGoal, describeScene,} and \textit{identifySubject} did not reach particularly low averages (all means equalling 1.8, 1.79, and 1.88 respectively, and SD=0.91, 0.83, 0.79, respectively), confirming the application’s mediocre performance.

\begin{figure}
\includegraphics[width=0.45\textwidth]{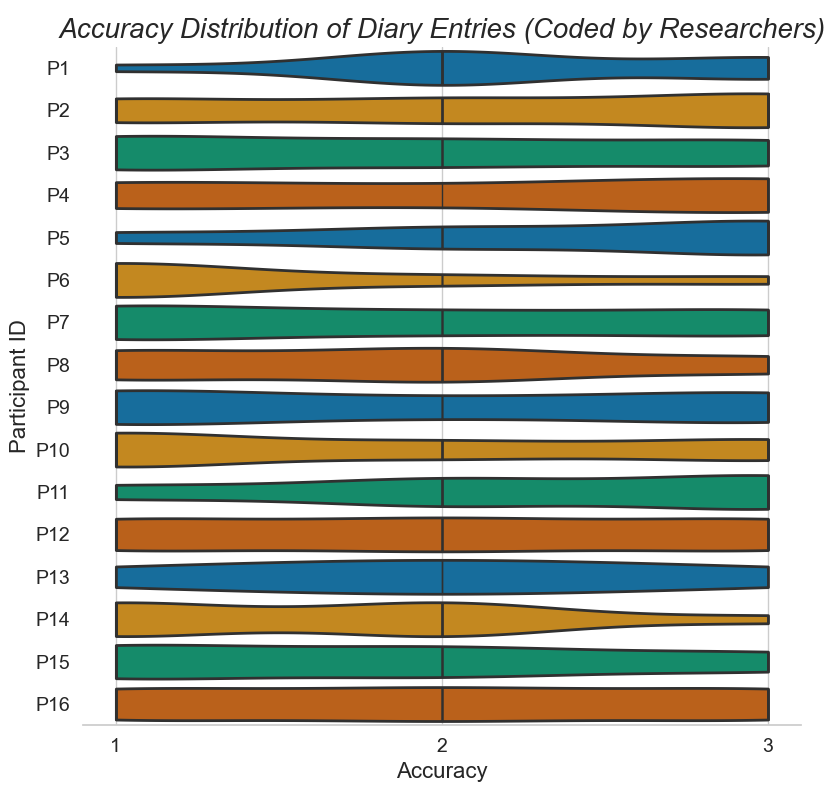}
\caption{\label{fig: violinplotsacc} The distribution of accuracy scores for descriptions in participants diary entries. These scores were assigned by two researchers looking at diary entries, photos, and descriptions to understand how well the application performed given the source material (see section \ref{Qualitative Analysis} for further detail).
}
\Description{Violin plot displaying participants' accuracy scores distribution across their diary entries. These accuracy scores were assigned by authors for each pair of image and generated description to get a general sense for the quality of descriptions participants received. P1 and P2 scores distributions for accuracy are evenly distributed, slightly on the higher end. P3, P4, P9, P10 and P15 scores distributions for accuracy are all relatively evenly distributed; P3, and P15 having slightly more diary entries with higher values. P6, P7, P8, P11, P13, and P14 scores distribution for accuracy are evenly distributed. P5, and  P11 score accuracy distributions had a higher proportion of diary entries with higher scores (e.g., better accuracy).
}
\end{figure}

\subsubsection{Trust and Satisfaction for Different Use Cases}

We quantified how useful the application was for identified use cases by looking at the impact of two major factors (goals and location) that would be most insightful towards understanding future use of AI-powered scene description applications. We also account for the accuracy of each diary entry, to understand whether it had an effect on participants’ satisfaction and trust. We used participants’ self-reported satisfaction and trust as predictors for future use of the application (i.e., higher trust and satisfaction means more use for the given purposes or locations). We interpreted “significance” as a p value < 0.05, established prior to running tests.

\begin{figure}
\includegraphics[width=0.45\textwidth]{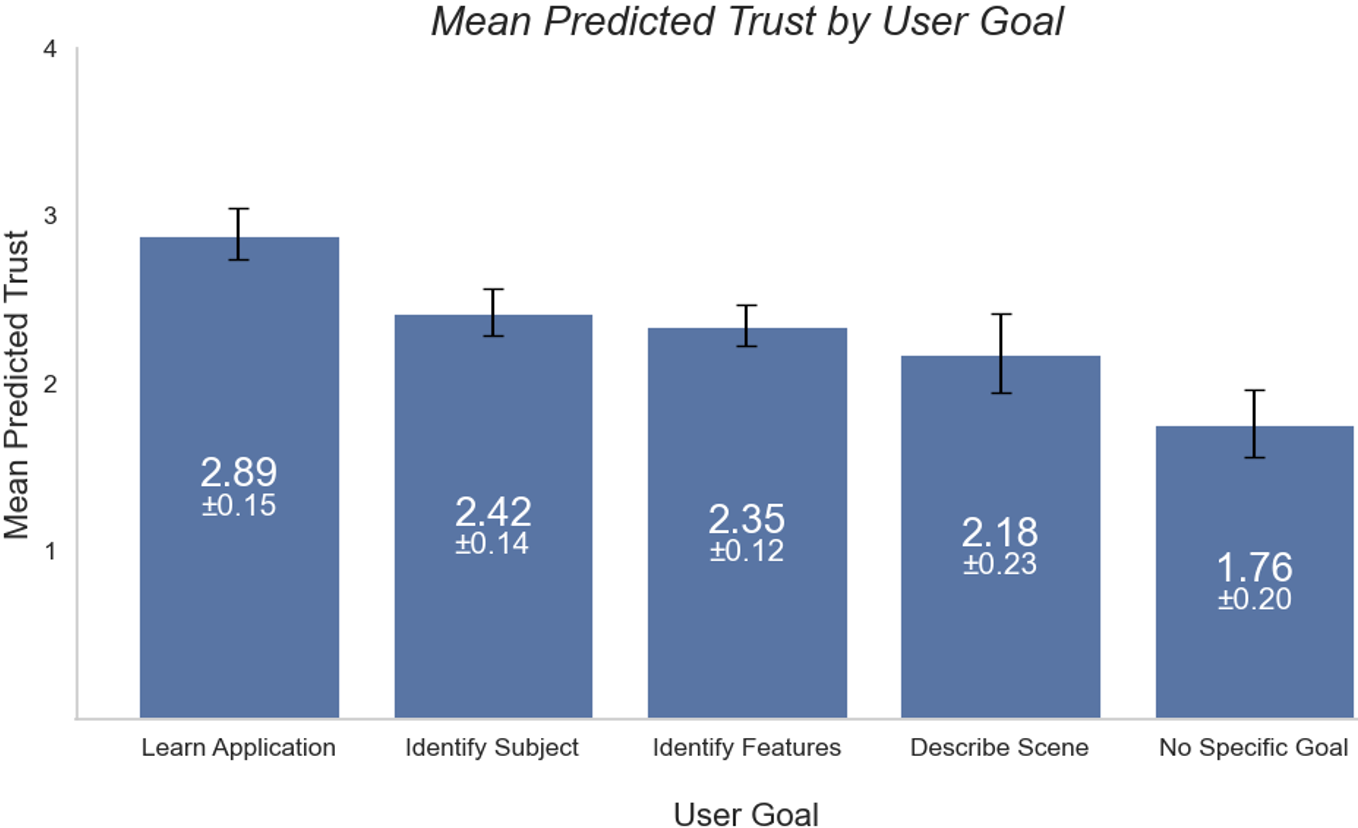}
\caption{\label{fig: predictedtrustgoal} The mean predicted trust per user goal estimated by our model.
}
\Description{Mean predicted trust per user goal estimated by our model. Learn Application user goal has the highest predicted score with 2.89 out of 4 with a standard deviation of 0.15. No Specific Goal has the lowest predicted score with 1.76 out 4 with a standard deviation of 0.2. Other user goals scores are: Identify Subject, Identify Features, and Describe Scene have predicted scores of 2.42 with standard deviation 0.14, 2.35 with  standard deviation 0.12, and 2.18 with standard deviation 0.23.
}
\end{figure}

\textbf{\textit{User Goals and Trust.}} We evaluated whether participants’ goals impacted their level of trust in  the description. We found a significant effect of \textit{User Goal} on \textit{Trust} (t=2.932, p<0.01, p=0.00374), and confirmed that interpretation accuracy had a significant effect on participants’ trust (t=11.386, p<0.001, p<2e-16). As with satisfaction, the trust a participant has for the application’s output will be significantly affected by that output’s accuracy. The goal \textit{learnApplication} also had a significant effect on trust (t=2.740, p<0.01, p=0.00654); pairwise comparisons found \textit{learnApplication} had a near significant difference on trust compared to \textit{describeScene} (p=0.0508), and a significant effect on \textit{identifyFeatures} (p=0.0057) and \textit{noSpecificGoal} (p<.0001). The mean trust participants gave while using the application with this goal was 2.89 (SE=0.150) out of 4. We conclude participants’ trust in the application was significantly higher when using it for testing or learning. No other goals had significant effects on trust.

\begin{figure}
\includegraphics[width=0.45\textwidth]{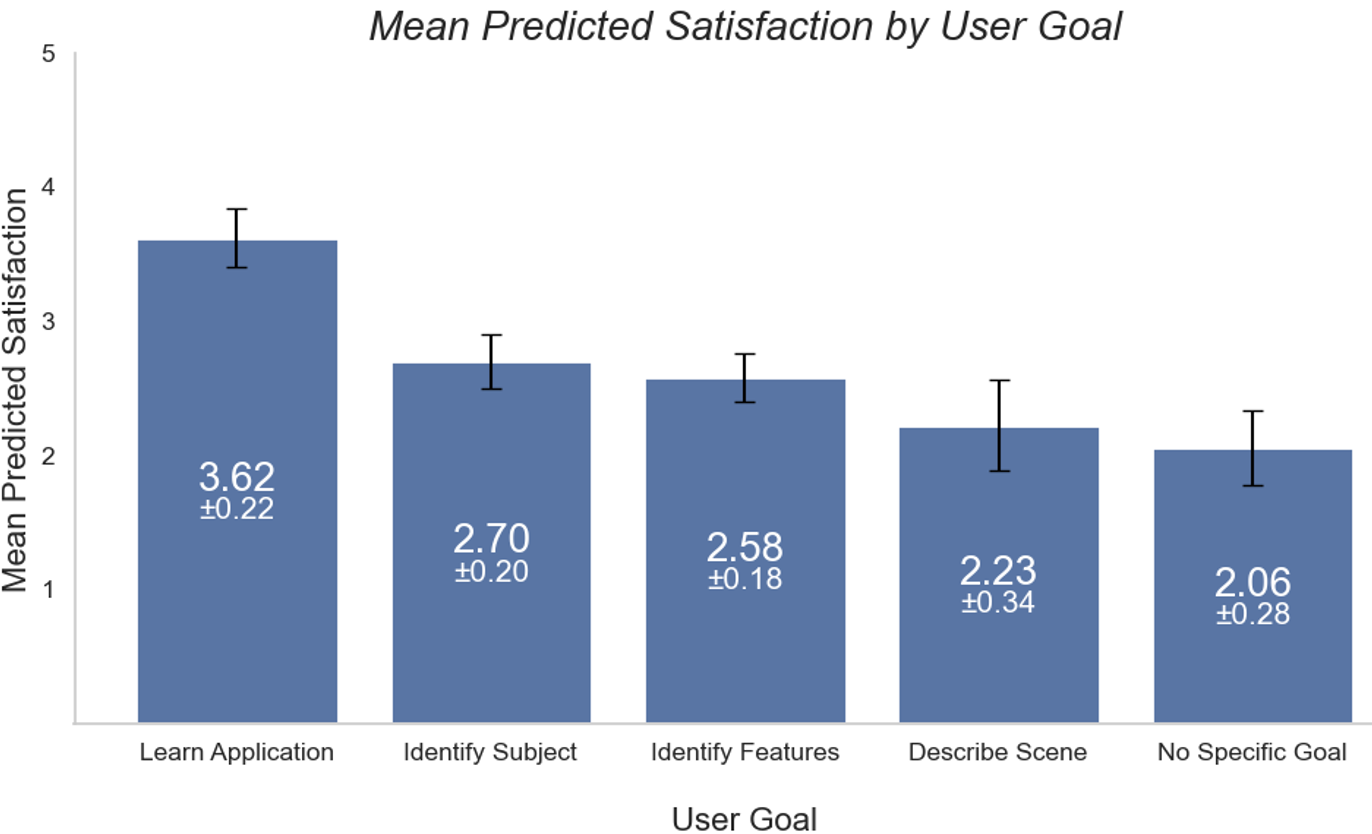}
\caption{\label{fig: predictedsatisfactiongoal} The mean predicted satisfaction per user goal estimated by our model.
}
\Description{Mean predicted satisfaction per user goal estimated by our model. Learn Application user goal has the highest predicted score with 3.62 out of 5 with a standard deviation of 0.22. No Specific Goal has the lowest predicted score with 2.06 out 5 with a standard deviation of 0.28. Other user goals scores are: Identify Subject, Identify Features, and Describe Scene have predicted scores of 2.70 with standard deviation 0.2, 2.58 with  standard deviation 0.18, and 2.23 with standard deviation 0.34.
}
\end{figure}

\textbf{\textit{User Goals and Satisfaction.}} We evaluated whether participants’ goals impacted their satisfaction in the description. We found a significant effect of \textit{User Goal} on \textit{Satisfaction} (t=2.240, p<0.05, p=0.026220), and confirmed that scene description accuracy had a significant effect on participants’ satisfaction (t=9.445, p<0.001, p<2e-16). As might be expected, we find interpretation accuracy will significantly affect a participant’s satisfaction with the application and willingness to use it. The goal \textit{learnApplication} also had a significant effect on satisfaction (t=3.667, p<0.001, p=0.000291); pairwise comparisons found \textit{learnApplication} had a significant difference on satisfaction compared to \textit{describeScene} (p=0.0027), \textit{identifyFeatures} (p<.0001), \textit{identifySubject} (p=0.0135), and \textit{noSpecificGoal} (p<.0001). The mean satisfaction participants scored when using the application for this goal was 3.62 (SE=0.22) out of 5. We conclude participants’ satisfaction with the application was significantly higher when they were using it for testing or learning about its limits. No other goals had significant effects on satisfaction.

\textbf{\textit{Location and Trust.}} We evaluated whether participants’ locations impacted their level of trust in the descriptions. We found a significant effect of  \textit{Location} on \textit{Trust} (t=3.554, p<0.01, p=0.000451), and confirmed that accuracy had a significant effect on trust (t=11.341, p<0.001, p<2e-16). As before, this indicates interpretation accuracy will significantly affect participants’ trust. As before with the satisfaction factor, no locations were found to have a significant effect on satisfaction. In addition, pairwise comparisons found no significant differences on trust of any locations compared to the others. The closest comparison was \textit{usedInRetailBusinesses} compared to \textit{usedInLivingSpacesOrDomiciles} (p=0.1731). The mean trust reported at retail businesses was 3.38 (SE=0.395) out of 4; when at living spaces or domiciles, the mean was 2.37 (SE=0.106). This suggests participants’ trust might have been higher when using the application at retail businesses as compared to living spaces, but not to a significantly different extent. Further, participants reported no significantly different trust when using the application in retail businesses over any other locations. No other locations had significant effects on trust.

\begin{figure}
\includegraphics[width=0.45\textwidth]{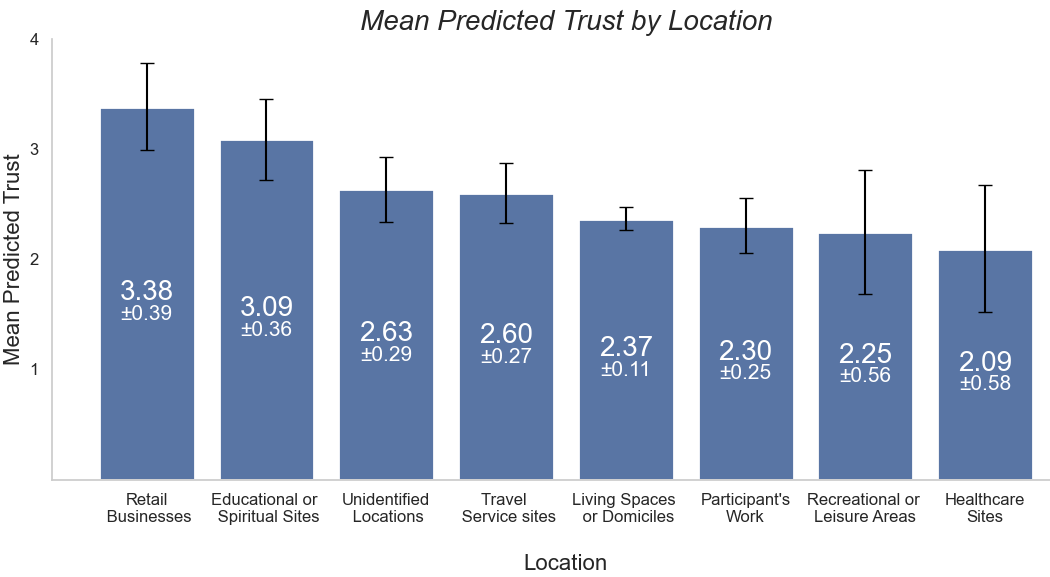}
\caption{\label{fig: predictedtrustlocation}The mean predicted trust per location estimated by our model.
}
\Description{Mean predicted trust per location estimated by our model. Retail Businesses location has the highest predicted score with 3.38 out of 4 with a standard deviation of 0.49. Healthcare Sites has the lowest predicted score with 2.09 out 4 with a standard deviation of 0.68. Other locations score are: Educational or Spiritual Sites 3.09 with standard deviation 0.36, Unidentified Locations 2.63 with standard deviation 0.29, Travel Service Sites 2.60 with standard deviation 0.27, Living Spaces or Domiciles 2.37 with standard deviation 0.11, Participant’s Work 2.30 with standard deviation 0.25, Recreational and Leisure areas 2.25 with standard deviation 0.56.
}
\end{figure}
\begin{figure}
\includegraphics[width=0.45\textwidth]{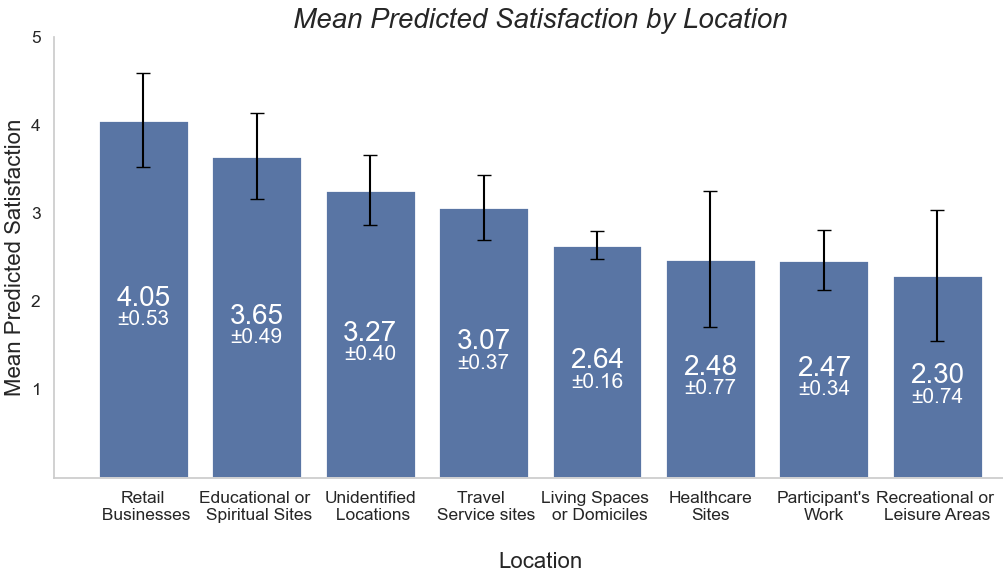}
\caption{\label{fig: predictedsatisfactionlocation} The mean predicted satisfaction per location estimated by our model.
}
\Description{Mean predicted satisfaction per location estimated by our model. Retail businesses have the highest predicted score with 4.05 out of 5 with a standard deviation of 0.53. Recreational and Leisure areas has the lowest predicted score with 2.30 out 5 with a standard deviation of 0.74. Other locations score are:  Educational and spiritual sites 3.65 with standard deviation 0.49, Unidentified Locations 3.27 with standard deviation 0.49, Travel Service Sites 3.07 with standard deviation 0.37,  Living Spaces or Domiciles 2.64 with standard deviation 0.16, Healthcare Sites 2.48 with standard deviation 0.77 and Participant’s Work 2.47 with standard deviation 0.34.
}
\end{figure}

\textbf{\textit{Location and Satisfaction.}} We evaluated whether participants’ locations impacted their satisfaction in the descriptions. We found a significant effect of \textit{Location} on \textit{Satisfaction} (t=3.798, p<0.05, p=0.000184), and confirmed that interpretation accuracy had a significant effect on participants’ satisfaction (t=9.370, p<0.001, p<2e-16). As with the model that included the user goal factor, interpretation accuracy remains a factor that significantly affects participants’ satisfaction and willingness to use the application. However, no locations were found to have had a significant effect on satisfaction. In addition, pairwise comparisons found no significant differences on satisfaction of any locations compared to the others. The closest comparison was \textit{usedInRetailBusinesses} compared to \textit{usedInLivingSpacesOrDomiciles} (p=0.1351). The mean satisfaction reported at retail businesses was 4.05 (SE=0.531) out of 5; when at living spaces or domiciles, the mean was 2.64 (SE=0.156). Thus, participants’ satisfaction might have been higher when using the application at retail business as compared to living spaces, but not to a significantly different extent. Further, participants reported no significantly different satisfaction when using the application in retail businesses over any other locations. No other locations had significant effects on satisfaction.

\subsubsection{Participant Definitions of Trust and Satisfaction} \label{definitions-of-trust-and-satisfaction}

\paragraph{\textbf{\textit{Defining Trust.}}} We held conversations with participants about the trustworthiness of the application’s descriptions, how trust was established, and how it could shift to compensate for inaccuracies. Several participants said their trust in the application developed over use. They alluded to a concept we called “practice ground of visual landscapes” (e.g., their home, neighborhood, nearby stores). These practice grounds would be used to test the AI against knowledge previously acquired from touch, residual vision, experience or past sighted assistance. Based on the application’s correctness over time, participants calibrated their trust in it and learned its capabilities. P15 gave a detailed explanation of her testing process:

\begin{myquote}
“[In many instances,] I knew what was in the frame for one reason or another whether it was using my residual vision or just previous experience. A lot of the photos were taken around my neighborhood and I have lived here for six years. Or if I happen to be out with my spouse, and I borrowed his eyeballs to verify something, you know, I am in the grocery store. Like the places where we shop all the time. I know where the sections of the stores I was standing in, for example. So, I had a very solid sense of the accuracy and the reliability and the usefulness of the information that was there.”

(P15, female, age 47)
\end{myquote}

Participants also shed light on seemingly paradoxical entries where descriptions with low accuracy scores received high trust and satisfaction scores. For example, P1 obtained the following interpretation while photographing a church pew: “A wood cabinet with a white device and a white device on it.” In her entry, she stated she wanted to know what was on the pew, and scored the interpretation unexpectedly high for satisfaction and trust (4 and 3). During her interview, she explained this was because she had figured out there were sermon notes: “it tells me about the wood, technically a shelf [...] Even though it's not a device, the fact that it said it was a white device [...I knew ] it was white. So based on that and the fact that there was that wood cabinet or shelf there, it kind of helps me deduce in my mind like, oh, it probably means a white piece of paper.”

We conclude trust in scene descriptions involves going beyond a correct versus incorrect binary judgment of the information provided. Even when interpretations were inaccurate, participants could still garner useful information if the inaccuracies were tolerable, maintaining relative trust in future interpretations.

\textbf{\textit{Defining Satisfaction.}} Participants defined satisfaction with various degrees of tolerance for inaccuracies, also noting that “accurate” results might not necessarily lead to satisfactory experiences. Most participants agreed useful information was necessary for high satisfaction. Definitions of useful, however, differed greatly. On one hand, some advocated that an application’s design should minimize work for the user. They explained the application should attempt to get things “right” on the first try by providing as much information as possible (P10, P11). P5 emphasized that “it should provide everything it saw from top to bottom, from best quality to questionable quality [...] In one instance [I might want] the background and in another instance it might be a facial expression. Who knows? How would it know what I want?” Others believed interpretations only needed to provide enough information to identify an object. Similar to the tolerable inaccuracies discussed earlier, this meant an interpretation would be satisfying as long as it gave enough details to support the context the participant already had, regardless of overall accuracy (P7).

In short, participants’ satisfaction was connected to the application’s performance, but not fully dependent on total correctness. Some participants would be satisfied by relatively poor scene descriptions, so long as they provided a few workable details. While participants preferred varying levels of detail, it seemed satisfaction overall was tied to the application’s ability to support their needs on a case-by-case basis, with the understanding that it could not perfectly interpret those needs on its own.
\section{Discussion}

To our knowledge, our work presents the first attempt to understand BLV people’s use cases for AI-powered scene description applications in their daily lives. Our findings revealed several novel uses for these applications, teasing out user behaviors with testing applications and triangulating known information with desired details. Our statistical analysis also demonstrated the significant effect of accuracy on users’ trust and satisfaction, as well as how these scores were significantly higher when testing the application. 
Finally, discussion with our participants revealed new perspectives on evaluating scene descriptions, such as matching the level of detail to users’ needs on a case-by-case basis and accepting tolerable inaccuracies, which supports prior work on BLV users’ acceptance of errors in assistive tools \cite{abdolrahmani2017embracing}.

\subsection{Leveraging BLV Users Abilities in Application Design}

The inherent design of many scene description applications is non-conversational (e.g., our application, SeeingAI \cite{seeingAI}, ImageExplorer \cite{image-explorer-guo}, ImageAssist \cite{nair-imageassist}), reflecting lack of awareness that BLV users are capable of verifying the information they receive, and can also gather information beforehand to provide as an input. Even when users can respond to initial descriptions with questions, they are unable to provide this information either before or along with the first photo submission in these applications. Prior work examines how these applications can better fulfill BLV users’ information needs but does not acknowledge that BLV users can and will gain information without the application’s assistance \cite{sankarnarayanantraining, Alharbi-Obfuscation-VDS}. However, research in negotiating social spaces has established that BLV people triangulate their information needs by collecting data from various non-visual sources \cite{thieme-everythingbutsee}. In our study, participants described many data collection methods they utilized prior to requesting scene descriptions, and they also revealed they knew about their photo subject by nature of the questions and answers given in the diary entries. BLV users apply these methods to validate or evaluate application results, rather than having nothing to rely on but the application. In fact, we noted our participants would often be irritated with the study application if it only provided information they already knew or could easily find themselves. This could inform the design of future applications to work with users’ existing knowledge and supply the information they actually need.

To illustrate this point, we propose that future applications tailor their output through settings and instruction from the user. This could include offering adjustments for the level of detail output, starting with generics and moving into specifics, and allowing users to tactilely explore photo “layers” for specific information. Such techniques for exploring digital content have been examined previously with screen readers and image exploration studies, showing promising results in providing a better understanding of which objects are present in the scene \cite{morris-rich-representation, nair-imageassist, image-explorer-guo}. By considering such examples, we can ensure visual interpretation applications adequately address the needs of BLV users. We additionally found that images could be perceivable in one dimension (e.g., clarity) while failing in another (e.g., lighting) due to complex image quality. Future systems could provide nuanced instructions to BLV users for improving photo capture to increase the quality of the input the AI receives.

\subsection{Visual Challenges In BLV People’s Daily Life}

Our study corroborated many visual challenges found in Brady et al.’s taxonomy \cite{brady-visual-challenges-everyday-lives}. Participants used the AI-powered application in similar ways to users of VizWiz Social, but we also revealed a few novel use cases. We found many examples for each of the challenges addressed in Brady et al.'s taxonomy in our diary entries: identifying objects (e.g., identity, media, currency), describing things (e.g., state, appearance, color), and reading (e.g., text labels, numbers). We also presented new categories of challenges that did not fit under Brady et al.’s taxonomy in the form of identified user goals, like building an understanding of a scene and taking tester photos to update one's understanding of an AI. In our study, 20\% (normalized) of submitted entries fell into these two categories. BLV users may be interested in identifying subtle things like the weather outside their window, receiving aesthetic descriptions of a landscape on vacation, or evaluating the AI’s effectiveness in new scenarios. Future applications in this domain should select environmental information to describe to a BLV audience, especially visual information that is difficult to triangulate through other means, as well as encourage frequent testing to increase awareness of improvements and create spaces for users to provide usage feedback (see Section \ref{Limitations and Future Work}).

We also uncovered new visual challenges that could be classified under Brady et al.’s taxonomy, like determining a subject's presence and location (under identification), or describing an object’s cleanliness and potential danger (under description). We believe these and many other unique user goals in our study became apparent due to the application being AI-powered. BLV users knew that no human (except for the researchers) would see the photos they submitted. With this awareness, they felt comfortable using the application in situations where requesting help from a remote assistant would be inappropriate or unnecessary (see Section \ref{Differentiating AI Use Cases from Human}). Noting this behavior, we believe AI-powered applications present unique opportunities to address visual challenges in BLV people's daily lives, not only for their scalability, but because even when human interpreters are available, AI might remain a preferable option to BLV users.

\subsection{Differentiating AI Use Cases from Human}\label{Differentiating AI Use Cases from Human}

Our study speaks to an existing body of work on human visual interpretation that has identified many use cases BLV people have for visual interpretation from human assistants. Lee et al. \cite{lee2018conversations, lee2020emerging} found four categories of visual interpretation support that BLV people seek: scene description, navigation, task performance, and social engagement. These categories held true in our findings, and reflected higher-level categories that many of the use cases we identified fall into. For example, our user goals “identify specific subject” and “determine subject state or amount” both involve gathering information about a subject in order to act on that information, like learning if a battery light is on so one can adjust a charger if needed, or learning if one is holding the correct shirt before wearing it. Such goals fall under Lee et al.’s “task performance” category and corroborate their findings on the kinds of support visual interpretation offers BLV people. However, we have also noted several new use cases, unique to AI-powered applications, that stand apart from the existing canon of use cases for human assistants.

We found our participants favored using AI over humans to minimize social costs, especially for questions they deemed not worthy of asking human helpers. For example, participants used the AI-powered study application for seemingly small questions, like differentiating sunglasses from prescription glasses or ensuring their sleeping roommate was not on the bed, because they did not want to bother other people. Participants felt using AI saved them the trouble of “burdening” their social network by requesting answers to such questions they considered to be trivial. To support this, future AI applications could remind users of their role as an alternative to social networks, for example, by being integrated as a “contact” in the user’s phone that they can message for quick descriptions before contacting family members or friends. Another AI-specific use case we found was using the AI as an “unbiased” information source to solve visual information disputes. For example, one participant and his BLV friend perceived the AI as an objective information source, and so in one instance they used the application as an arbiter to decide whether he or his friend was correct about the view outside their airplane window. Future AI applications could include disclaimers about their description accuracy to clarify how valuable their "unbiased" opinions should be in such situations.

Finally, one important AI-specific use case was testing the application, which was a goal applied to 12\% of the photos in our dataset. Based on our discussions with participants, a BLV user would likely consider it burdensome to the human assistant if they submitted photos merely to challenge the human assistant or out of curiosity about their responses. Yet such testing is vital for users to comprehend an AI's functionality, optimize its use, and see how it has improved over time. Given the rapid evolution of contemporary AI algorithms, continuous testing will remain crucial for users to stay informed about the AI's capabilities. In addition, whenever users encounter situations where they have not used the AI, or it has failed them before, they will likely test the application to see how it performs.

In conclusion, our discussions with participants highlighted significant differences between using AI-powered systems and human assistance. Trade-offs in privacy, independence, and social costs exist when working with humans and can be higher when compared to when working with AI. Conversely, despite AI advancements, there are trade-offs in the accuracy, precision, understandability, and reliability of AI systems, which may differ from trade-offs associated with leveraging human assistance. Acknowledging these distinctions, we assert that assumptions about how BLV individuals use human-powered systems cannot perfectly explain how they use AI-powered systems. This work marks an initial step in establishing a distinct role for AI-powered description, examining it separately from human-powered visual interpretation, and seeks to understand the differences between these roles.

\section{Limitations \& Future Work} \label{Limitations and Future Work}
This research was completed over the course of two years, during which the capabilities of AI improved dramatically. When we conducted the diary study, LLM and VQA experiences fine-tuned for image description, like BeMyAI, were unavailable \cite{be-my-ai}. These advancements will affect the accuracy and the user experience of AI-powered scene description applications, and future work should continue to investigate the ways in which use cases and user assessments of descriptions evolve. We hope this paper will serve as a useful point of comparison. 

We also acknowledge choosing a diary study to collect data meant we accepted trade-offs for collecting real-world data at the cost of inconsistent numbers of entries from participants. Moreover, participants’ submissions were strongly tied to their seasonal schedules. For example, if they were vacationing, photos happened in locations not necessarily representative of their “daily lives”, such as resorts in a foreign country. If a participant worked from home during the study period, most photos were taken indoors in living spaces. We recommend future studies integrate notifications to remind users to upload submissions throughout the study at different times and locations.

Finally, a significant amount of entries fell into the “Testing Application Against Current Knowledge” use case, which may be an artifact of the study conditions. It is possible that participants felt obliged to use the application but had no organic need, so they simply tested its capabilities. On the other hand, it is also possible that the need to test the application was organic, especially considering that they were all using the application for the first time when the study began. As participants encountered new situations (e.g., subjects of interest, locations, etc.), they wanted to know whether the application would help them address a new visual challenge. Future studies could be held for longer periods, and thus extend their participants’ exposure to the study application(s), so that testing new visual challenges will represent a smaller part of a dataset. Alternatively, they could implement applications that participants are already familiar with to lessen the anticipated testing period.

Based on the discussion and the limitations of our work, we propose two directions for future research in this area:

(1) Researchers should investigate whether integrating a wider variety of features  (e.g., bar code scanning, feature description, OCR, and exploring images by touch) in one holistic experience could reveal new use cases for AI-powered scene descriptions. In addition, researchers can also investigate use cases for generative AI-powered scene descriptions applications like BeMyAI following a similar procedure to ours.
(2) Researchers should investigate the design of the feedback loop Human-AI-Technology, how to better integrate user feedback based on their capabilities, and how to optimize the feedback loop. For example, a built-in testing feature could be integrated, notifying users of model updates, allowing them to provide feedback on use case results, and encouraging tests on familiar scenarios to grasp its updated behavior. The most important aspects about the feedback loop would be: feedback during use (e.g., communicating with the AI), feedback post-use (e.g., providing feedback about AI behavior and forms of communication), and asynchronous 
\section{Conclusion} \label{Conclusion}
We conducted a diary study with 16 BLV participants that has highlighted overlooked factors of BLV users’ engagement with scene description applications, such as their ability to gather prior knowledge about queries. We also revealed common use cases, such as determining the specific identity of an object, and the context in which BLV people use these applications through a categorization of user goals, locations, photo content, and photo quality. In addition, we have pointed out where AI-powered interpretations fall short in addressing these daily use cases and where they succeed, based on three evaluation metrics of trust, satisfaction, and accuracy. Lastly, we highlighted AI-specific use cases for scene description to be considered in future systems, such as using AI to minimize social costs of interpretation. 

Scene description applications and the AI powering them are changing rapidly, especially with continuing advancements in generative AI. As such technologies grow, it is critical to examine how users interact with these technologies, what their needs are, and what their goals for these technologies are. These investigations will illuminate advancements for visual interpretation that will be most useful to BLV users going forward. This study and the dataset we share can be used for investigating the effectiveness of generative models like GPT-4, similar to the evaluations conducted by Gurari et al. \cite{gurari2018vizwiz} on VizWiz. We hope our work and the data collected will guide advancements in scene description and establish a more useful baseline of visual interpretation for BLV users and their daily visual needs.

\bibliographystyle{ACM-Reference-Format}
\bibliography{main}

\appendix

\end{document}